\documentclass[aps,prb,nofootinbib,twocolumn,amsmath,amssymb,groupedaddress,floatfix]{revtex4-2}

\usepackage{amsthm}
\usepackage{mathtools}
\usepackage{mathdots}
\usepackage{amsfonts}
\usepackage{bbm}
\usepackage{bbold}
\usepackage{color,dsfont,upgreek}
\usepackage{mathrsfs}
\usepackage[hidelinks]{hyperref}
\usepackage[normalem]{ulem}

\usepackage{hyperref}
\hypersetup{colorlinks,
	linkcolor={blue!75!black!80!yellow},
	citecolor={blue!75!black!80!yellow},
	urlcolor={blue!75!black!80!yellow}
}

\usepackage{svg}
\usepackage{graphicx}

\usepackage{dcolumn} 
\usepackage{bm} 
\usepackage{float} 

\allowdisplaybreaks

\newcommand{\diff}{{\mathrm{d}}}
\newcommand{\ddt}{{\frac{\mathrm{d}}{\mathrm{d}t}}}

\newcommand{\im}{{\mathrm{i}}}

\begin{document}

\title{Solitons with self-induced topological nonreciprocity}
\author{Pedro Fittipaldi de Castro}
\author{Wladimir A. Benalcazar}
\email{benalcazar@emory.edu}
\affiliation{Department  of  Physics,  Emory  University, Atlanta, Georgia 30322, USA}

\begin{abstract}
The nonlinear Schr\"odinger equation supports solitons---self-interacting, localized states that behave as nearly independent objects. Here, we show the existence of solitons with self-induced nonreciprocal dynamics in a discrete nonlinear Schr\"odinger equation. This nonreciprocal behavior, dependent on soliton power and symmetry, occurs when parity is broken in a lattice with an Ablowitz-Ladik type of nonlinearity. Initially stable at high power, solitons exhibit nonreciprocal instabilities as power decreases, leading to unidirectional acceleration and amplification. This behavior is topologically protected by winding numbers on the solitons’ mean-field Hamiltonian and their stability matrix, linking nonlinear dynamics and point gap topology in non-Hermitian Hamiltonians.
\end{abstract}

\date{\today}
\maketitle
\noindent

Nonreciprocity, or the unequal behavior of channels moving in opposite directions, is a peculiar feature of physical systems that lack spatial reflection and time-reversal symmetries. It has wide applications in optical and electrical circuit components~\cite{sounas2017non,nagulu2020non}, such as isolators~\cite{jalas2013and} and circulators~\cite{tanaka1965active}, which are crucial for source protection and prevention of spurious interferences. Some platforms rely on a magnetic field to break the necessary symmetries~\cite{doi:https://doi.org/10.1002/0471213748.ch6,dotsch2005applications,shoji2012mzi,wang2005optical,bi2011chip}, while others achieve nonreciprocity, for example, by spatially and temporally modulating resonator arrays~\cite{kamal2011noiseless,yu2009complete,hafezi2012optomechanically,PhysRevLett.123.063901,kim2015non}.

When defined in a crystalline structure, some systems display a robust, topologically protected form of nonreciprocity that occurs, for instance, at the one-dimensional (1D) edges of systems hosting the quantum Hall effect~\cite{PhysRevLett.67.749,PhysRevLett.64.220,PhysRevLett.71.3697,PhysRevLett.61.2015} and in non-Hermitian lattices with a ``skin effect''~\cite{PhysRevB.97.121401,rotter2007nonhermitian,PhysRevLett.123.170401,PhysRevB.99.125155,PhysRevLett.121.026808}. In the former case, edge states exhibit unidirectional propagation due to a topological bulk-boundary correspondence~\cite{RevModPhys.95.011002}, while in the latter, the energy spectra take complex values $E_n(k) \in \mathbb{C}$ that wind around a certain point in the complex energy plane across the Brillouin zone~\cite{PhysRevLett.121.086803,PhysRevX.8.031079,knotsNonHermitian}. The resulting winding number is a topological invariant associated with fields that exhibit unidirectional amplification and acceleration~\cite{PhysRevB.105.245143,PhysRevLett.123.170401,PhysRevB.108.125402,PhysRevB.105.165137}.

In parallel with the development of non-Hermitian quantum mechanics, advances in condensed matter, such as the recent observation of fractional Chern insulators~\cite{Regnault_2011, Sheng_2011, PhysRevLett.106.236804, PhysRevLett.107.146803}, have highlighted the importance of interactions in enabling novel topological phases. Still, the role of interactions and nonlinearities as catalysts of topologically protected nonreciprocity remains largely unexplored. One way of attacking the problem is to consider nonlinear lattices supporting discrete solitons or breathers---collective self-interacting localized states that behave as individual particles, with well-defined quantized properties~\cite{RevModPhys.83.247,LEDERER20081}---and analyze their dynamical properties. 

In this Letter, we break parity in the Ablowitz-Ladik (AL) model~\cite{10.1063/1.522558,10.1063/1.523009}, a discretization of the nonlinear Schr\"odinger equation, and show that it supports solitons with self-induced topological nonreciprocity. While the asymmetry in the spatial distribution of our solitons about their ``center of mass'' is necessary for their nonreciprocal dynamics, it is not sufficient. Additionally, their nonreciprocal behavior is topologically protected by point gaps with nontrivial windings in the solitons' (non-Hermitian) stability and mean-field (MF) energy spectra under periodic boundary conditions (PBCs), thus unveiling an intimate connection between nonlinear, nonreciprocal dynamics and point gap topology in non-Hermitian linear Hamiltonians. Indeed, as we shall see, the nonreciprocal characteristics of the solitons described herein are akin to those found in the wave packet dynamics of linear systems exhibiting a skin effect~\cite{PhysRevB.106.L241112,PhysRevB.105.245143}.

The type of nonlinearity we describe appears in the approximate descriptions of the dynamics of small-amplitude excitons in atomic chains with exchange or dipole-dipole interactions~\cite{PhysRevB.55.11342}, energy transport in biological systems~\cite{PhysRevE.61.5839}, or even solitons in nonlinear waveguides~\cite{PhysRevE.53.1172}.

\begin{figure}
    \centering
    \includegraphics[width=\columnwidth]{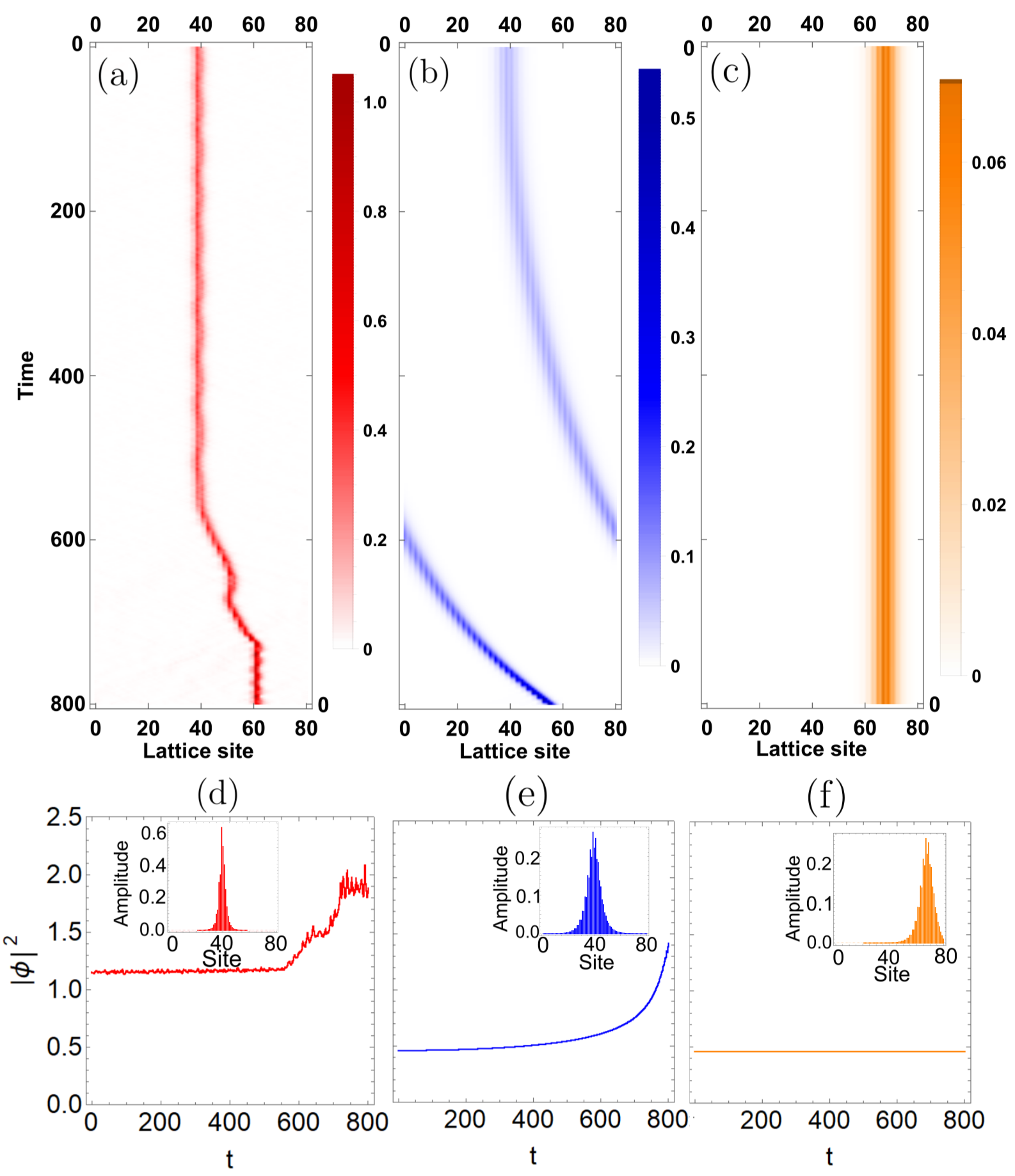}
    \caption{Dynamics of nonreciprocal solitons. (a)-(c) Time evolution of the square amplitude per site of solitons in the parity-broken AL model, Eq.~\eqref{OurModel}. In (a), the initial configuration is a static soliton with norm $|\phi|=1.05$, energy $\omega=-2.32$, period of oscillation $T=2\pi/|\omega| = 2.71$, and random complex perturbations $|\delta\phi_a|<0.0001$ added across the lattice at each time step under PBC. In (b), the initial configuration is an unperturbed static soliton with norm $|\phi|=0.68$, energy $\omega = -2.09$, and period of oscillation $T=2\pi/|\omega| = 3.01$ under PBC. (c) Evolution of the square amplitude of a soliton with $|\phi|=0.68$, energy $\omega = -2.09$, and period of oscillation $T=2\pi/|\omega| = 3.01$, but now located next to the boundary under open boundary conditions. (d)-(f) Field intensity $|\phi|^2$ as a function of time for the solitons displayed in (a)-(c). The insets in (d)-(f) show the field profiles $\phi_R$ at $t=0$.
    All simulations use the fourth-order Runge-Kutta algorithm with a time step $\Delta t = 0.01$ and lattice parameters $t_1=1.35$, $t_2=0.65$, $\Delta=0.35$, and $L=40$ unit cells ($80$ sites).}
    \label{DynamicSimulations}
\end{figure}

\emph{Model.} As a minimal example of a nonlinear system supporting self-induced nonreciprocal solitons,
we consider a field $\Phi$ defined in a 1D lattice and governed by the equations of motion
\begin{align}
    \im\ddt \Phi_a = -\sum_{b}(h_{ab}+|\Phi_a|^2 f_{ab})\Phi_{b}, \label{OurModel}
\end{align}
where the composite index $a = (R,\alpha)$ runs over unit cells $R=1,\dots,L$ and their internal degrees of freedom $\alpha=A,B$ (the lattice basis). The linear part of \eqref{OurModel} is determined by the tight-binding matrix
\begin{align}
    h_{(R,\alpha)(R',\beta)} &= \left(t_1\sigma^x_{\alpha\beta}+\Delta\sigma^z_{\alpha\beta}\right)\delta_{R,R'} \nonumber \\
    & \ \ +t_2\left(\sigma^{+}_{\alpha\beta}\delta_{R-1,R'}+\sigma^{-}_{\alpha\beta}\delta_{R+1,R'}\right), \label{tighbinding}
\end{align}
where $\sigma^{x},\sigma^{y},\sigma^{z}$ are the Pauli matrices and $\sigma^{\pm} = (\sigma^x \pm i \sigma^y)/2$. Here, $t_1$ and $t_2 >0$ are intra- and inter unit cell (nearest neighbor) NN hopping parameters, and $\Delta$ represents the staggered on-site energies of the $A$ and $B$ sublattices.

The nonlinear contribution to \eqref{OurModel} is proportional to
\begin{equation}
    f_{(R,\alpha)(R',\beta)} = (\sigma^{x}_{\alpha\beta}\delta_{R,R'}+\sigma^{+}_{\alpha\beta}\delta_{R-1,R'}+\sigma^{-}_{\alpha\beta}\delta_{R+1,R'}), \label{tighbindingnonlinear}
\end{equation}
and also connects NNs. The probability density function (PDF) $|\Phi_{R,\alpha}|^2$ controls the strength of the nonlinear hopping from the site $(R,\alpha)$ to its NN destination.
The system breaks parity symmetry by setting $t_1 \neq t_2$ and $\Delta \neq 0$, which is a necessary condition for nonreciprocal solitons. Still, their nonreciprocal dynamics additionally depend on the power of the soliton's field. At high enough powers, parity-broken solitons are always static (and hence reciprocal). At lower powers, nonreciprocal linear instabilities appear on static solitons [Fig.~\ref{DynamicSimulations}(a)], while at even lower powers, the static soliton solutions disappear, replaced instead by solitons with sustained unidirectional acceleration and amplification [Fig.~\ref{DynamicSimulations}(b)], except at the one end of the lattice to which the nonreciprocal solitons accelerate, where a single static soliton solution persists [Fig. \ref{DynamicSimulations}(c)].

\emph{Static solitons and their linear stability.}
To expose the nonreciprocal behavior in our model, we investigate the fate of static solitons in the bulk of a lattice obeying \eqref{OurModel} as we vary the parity-breaking parameters $t_2/t_1$ and $\Delta$. We find these solutions by changing variables to the rotating frame $\Phi_a(t)=e^{-\im \omega t}\phi_a(t)$, where $\omega \in \mathbb{R}$ is some real frequency to be determined~\cite{schindler2023nonlinear}. In the rotating frame, the equations of motion \eqref{OurModel} take the form
\begin{equation}
    \im\ddt \phi_a = -\sum_{b}\left(h_{ab}+\omega\delta_{ab}+|\phi_a|^2 f_{ab}\right)\phi_{b}. \label{RotatingFrame}
\end{equation}

The static solitons we are after must be stationary in the rotating frame, $\ddt \phi_a=0$~\cite{FLACH1998181}, yielding the time-independent expression
\begin{align}
    \omega \phi_a &= -\sum_{b}(h_{ab}+|\phi_a|^2 f_{ab})\phi_{b} \equiv \sum_b H_{ab}(\phi)\phi_{b}, \label{SelfConsistent}
\end{align}
where $H(\phi)$ is a mean-field (MF) Hamiltonian built using the state $\phi$ with frequency $\omega$. We self-consistently solve for $\phi$, $\omega$, and $H(\phi)$ as detailed in the Supplemental Material (see also Ref. \cite{Cuevas_Maraver_2019} therein).

In addition to $\phi$, $H(\phi)$ has a multitude of extended eigenvectors $\psi$ with energies $\varepsilon$. In the $|\phi|^2 \to 0$ limit, the mean-field Hamiltonian reduces to the tight-binding matrix $h$ in \eqref{OurModel}, so the eigenstates $\psi$ are simply Bloch waves forming the two energy bands
\begin{equation*}
    \varepsilon_{\pm}(k) = \pm \sqrt{(t_1+t_2\cos k)^2+(t_2\sin k)^2+\Delta^2}
\end{equation*}
of a linear Rice-Mele chain~\cite{PhysRevLett.49.1455}. The extreme levels of the linear dispersion are $E_{\pm} = \pm \sqrt{(t_1+t_2)^2+\Delta^2}$. As $|\phi|^2$ increases, the Bloch waves at the edges $E_{\pm}$ of the two energy bands continuously deform to give rise to two localized eigenvectors of $H(\phi)$: the soliton $\phi$ itself and a state $\phi'$ with frequency $-\omega$. Thus, we say that a given soliton ``bifurcates'' from the edge of a particular energy band. From now on, we will focus on the properties of the soliton $\phi$ bifurcating from the lower band, corresponding to the ground state of the MF Hamiltonian. Although the energy $\omega$ of the soliton is real, $H(\phi)$ is non-Hermitian; thus, the energies $\varepsilon$ of the extended states may be complex. From \eqref{SelfConsistent}, we see that $H^{*}(\phi)=H(\phi) \neq H^{T}(\phi)$. Therefore, the MF Hamiltonian has time-reversal symmetry (TRS) and breaks the ramified version of TRS for non-Hermitian Hamiltonians, called $\mathrm{TRS}^{\dagger}$, lying in the symmetry class AI of the 38-fold classification of non-Hermitian Hamiltonians~\cite{PhysRevX.9.041015}, which has a $\mathbb{Z}$ topological invariant in one spatial dimension in the presence of a point gap.

Along with its mean-field spectrum, we probe the soliton's linear stability by adding arbitrary time-dependent complex perturbations $\delta\phi(t)$ to them, so the field becomes $\tilde{\phi}_a(t)=\phi_a+\delta\phi_a(t)$. Without loss of generality, we write the complex perturbations as $\delta\phi_a = v_ae^{\im\lambda t}+w_a^*e^{-\im\lambda^* t}$~\cite{Discrete}. Substituting $\tilde{\phi}_a(t)$ into \eqref{RotatingFrame} and expanding up to linear order in $\delta\phi_a$, we obtain the linear equation for the fluctuations around the soliton~$\phi$
\begin{equation}
    \lambda 
    \begin{pmatrix}
        v \\
        w
    \end{pmatrix}
    =
    M(\phi)
    \begin{pmatrix}
        v \\
        w
    \end{pmatrix}
    , \quad 
    M(\phi) \equiv \begin{pmatrix}
        \mathcal{A} & \mathcal{B} \\
        -\mathcal{B}^* & - \mathcal{A}^*
    \end{pmatrix},
    \label{StabCondition}
\end{equation}
where $v = [v_{(1,A)},v_{(1,B)},\dots,v_{(L,A)},v_{(L,B)}]^T$ is the vector containing the component $\{v_a\}$ for all choices of $a$ in the lattice basis, and similarly for $w$. The $\mathcal{A}$ and $\mathcal{B}$ matrices forming the non-Hermitian stability matrix $M(\phi)$ are given in Eq. (18) of the Supplemental Material.
The linear stability of the soliton $\phi$ is determined by the spectrum of $M(\phi)$: If at least one $\lambda$ has a positive imaginary part, the corresponding perturbation $\delta\phi(t)$ grows exponentially as a function of $t$, indicating that the original solution is unstable. By construction, the stability matrix enjoys the ramified version of particle-hole symmetry for non-Hermitian systems (called $\mathrm{PHS}^{\dagger}$), $\Gamma M(\phi)^* \Gamma = -M(\phi)$, with
$\Gamma = [0 \; 1 ; 1 \; 0]$ and as a consequence, the eigenvalues of $M(\phi)$ must come in pairs $(\lambda,-\lambda^*)$ or lie in the imaginary axis. Also, since the tight-binding matrices $h$ and $f$ are real and we can choose $\phi_a \in \mathbb{R}$ without loss of generality for static solitons, we get $M(\phi)^*=M(\phi)$, which means that the eigenvalues must come in complex conjugate pairs $(\lambda,\lambda^{*})$ or lie on the real axis. Together with $\mathrm{PHS}^{\dagger}$, the reality of $M(\phi)$ implies that complex stability eigenvalues must generally come in quadruplets $(\lambda,\lambda^*,-\lambda,-\lambda^*)$. Therefore, if there is one $\lambda$ with a negative imaginary part, there will necessarily be other eigenvalues with positive imaginary parts. Therefore, we arrive at our criterion: A soliton is linearly stable if and only if the spectrum of the stability matrix $M(\phi)$ is purely real. Lastly, $\mathrm{PHS}^{\dagger}$ makes $M(\phi)$ fall into symmetry class $D^{\dagger}$ of the 38-fold classification, which is also characterized by a $\mathbb{Z}$ topological invariant in the presence of a point gap.

\begin{figure}
    \centering
    \includegraphics[width=\columnwidth]{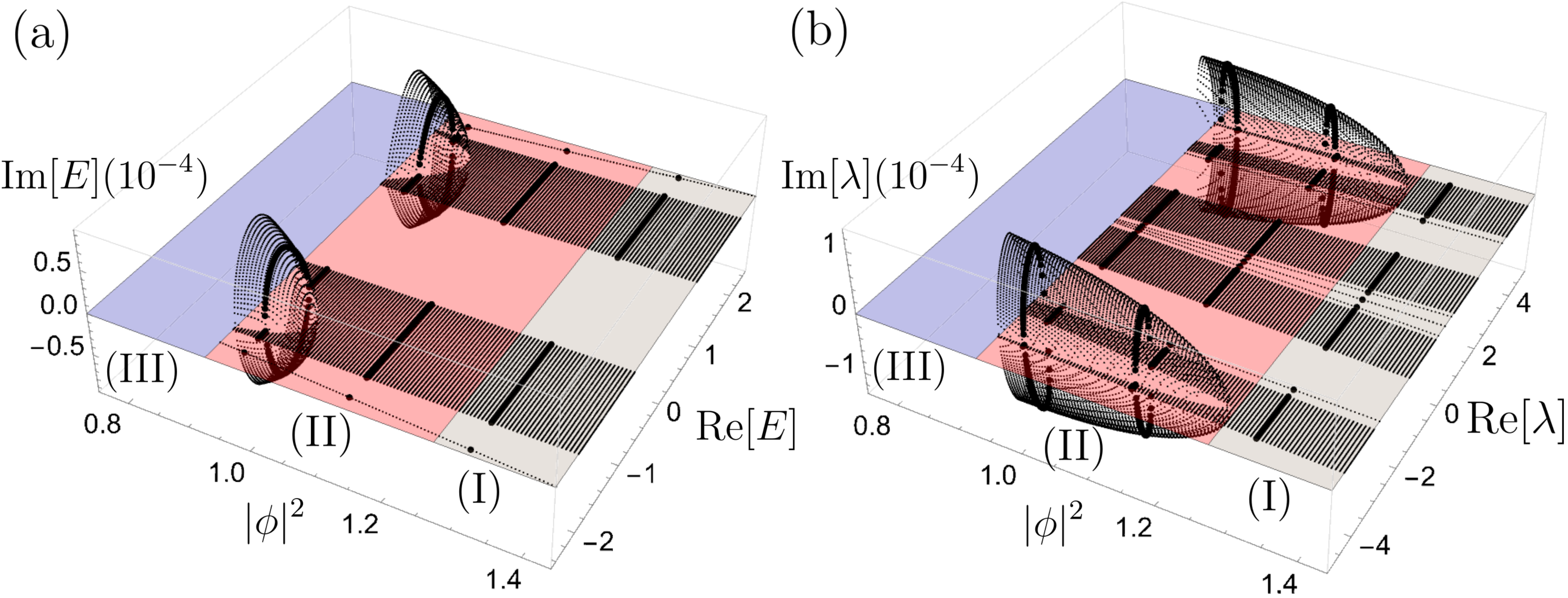}
    \caption{Complex mean-field and stability spectra of static solitons as a function of their norm squared $|\phi|^2$. (a) Eigenvalues $E$ (small black dots) of the mean-field Hamiltonian $H(\phi)$ showing two bands associated with extended states and two energies outside the gap, corresponding to the soliton $\phi$ (ground state) and its chiral partner. (b) Eigenvalues (small black dots) $\lambda$ of the stability matrix $M(\phi)$ stemming from the same solitons as in figure (a). Now, there are four bands because $M$ is twice the size of the mean-field Hamiltonian. In both (a) and (b), the gray, red, and blue regions correspond to regimes (I), (II), and (III), and the thick dots are the stability spectra of solitons at representative points [one in regime (I) and two in regime (II)]. Parameters used in the numerical calculations: $t_1=1.35$, $t_2=0.65$, $\Delta=0.35$, and $L=200$ unit cells.}
    \label{EnergyNonlinearity}
\end{figure}

\emph{Nonreciprocal effects.} The spectra of the MF and stability Hamiltonians, Eqs.~\eqref{SelfConsistent} and \eqref{StabCondition}, fall into three different regimes under PBC, illustrated in Fig.~\eqref{EnergyNonlinearity} (for the sake of simplicity, in the present discussion, we set the lattice parameters to $t_1=1.35$, $t_2=0.65$, and $\Delta=0.35$ for all simulations, and vary the intensity $|\phi|^2$ of the self-consistent solitons, which encode the nonlinearity strength). First, there is a strongly nonlinear regime (I) at $|\phi|^2 > 1.25$ supporting static bulk solitons that are linearly stable (all stability eigenvalues $\lambda$ are real) and whose mean-field Hamiltonians have purely real spectra consisting of two bands stemming from the extended eigenvectors $\psi$ and two localized states outside the gap, the soliton $\phi$ and its chiral partner, bifurcating from the lower and upper band, respectively.
Then, at $|\phi|^2=1.25$, some stability eigenvalues $\lambda$ become complex, opening \emph{point gaps} in the interiors of two out of the four stability bands. As we continue to reduce power, point gaps eventually open in intervals within both MF energy bands, but remain closed near the outermost edges $E_{\pm}$ as long as $|\phi|^2>0.94$. Thus, $0.94 < |\phi|^2 < 1.25$ defines an intermediate regime (II). When $|\phi|^2=0.94$, the point gaps are open at points in the real axis arbitrarily close to $E_{\pm}$ (see Sec. VI of the Supplemental Material for detailed plots of the MF spectrum close to this transition). Below this value, we obtain a weakly nonlinear domain (III) where the iterative sequence $\{\phi^{(i)}\}$ in the self-consistent mean-field solution stops converging to machine precision and instead reaches a lower bound $\delta \sim 10^{-3}$, suggesting the absence of \emph{stationary} solitons in that regime.

The complex stability spectrum indicates that the solitons in regime (II) are linearly unstable and that two of the stability bands exhibit point gaps in their interior. Our Supplemental Material demonstrates that both point gaps always have the same winding number $\nu$, which could be $\nu=+1$ or $\nu=-1$. We evaluate $\nu$ employing the method from Refs.~\cite{PhysRevX.9.041015,PhysRevB.105.165137} for determining the winding number of U(1)-symmetric systems without translation invariance, which involves the deformation of the stability matrix $M(\phi)$ as $M_{ab}(\phi) \mapsto e^{\im\theta(a-b)/2L}M_{ab}(\phi)$. Then, the quantity $\det[M_{\theta}(\phi)-\lambda_0]$ is periodic in $\theta$ and the winding of the stability matrix relative to a point $\lambda_0 \in \mathbb{C}$ is given by
\begin{equation}
\nu(\lambda_0) \equiv \frac{1}{2\pi\im} \int_{0}^{2\pi} \diff \theta \ \partial_{\theta} \ln \det \left[M_{\theta}(\phi)-\lambda_0\right] \;\; \in \mathbb{Z}, \label{winding}
\end{equation}
provided that $\det[M_{\theta}(\phi)-\lambda_0] \neq 0$ for all $\theta \in (0,2\pi)$. We can analogously define the winding of the MF Hamiltonian $H(\phi)$, as shown in Eq.~(27) of the Supplemental Material. Whenever the point gaps in the energy spectrum are open, we also observe that $H(\phi)$ has a winding number $\nu = \pm 1$.

Since the winding number is odd under parity, the overall effect of growth and decay of perturbations must favor motion in a particular direction, as in non-Hermitian systems with a skin effect. Figure \ref{DynamicSimulations}(a) shows the time evolution of an initially stationary soliton in regime (II) in the presence of random perturbations $\delta\phi_a$, confirming our prediction that the perturbed soliton exhibits an overall displacement in the direction determined by the winding number (right direction for $\nu=+1$ and left for $\nu=-1$) of the stability bands and the mean field spectrum below $|\phi|^2<1.25$.~\footnote{Our simulations of the evolution of solitons $\phi$ plus perturbations $\delta\phi$ using the RK4 method agree with the first-order transition value obtained from linear stability theory, and only start deviating from it for $|\phi|>0.18$ (see Fig. 1 in the Supplemental Material).}. In regime (II), the field intensity $|\phi|^2$ is no longer constant over time, in contrast to solitons in the usual AL model~\cite{wang2020dynamics,xie2017soliton}. Figure \ref{DynamicSimulations}(d) shows that the intensity of the unstable soliton grows when it drifts in the preferred direction.

The most extreme behavior occurs in the weakly nonlinear regime (III). Although not static, the resulting states have essentially the same form as the solitons in the previous regimes. Under time evolution, they remain localized, but now smoothly accelerate and amplify as they move in the direction determined by the winding number [Figs.~\ref{DynamicSimulations}(b) and \ref{DynamicSimulations}(e)]. Such a phenomenon is consistent with the fact that, at low powers, the soliton $\phi$ inherits the properties of the extended modes $\psi$ with energies near the band edge $E_{-}$ from which it bifurcates~\cite{jurgensen2023quantized,schindler2023nonlinear}.

A regime having no static bulk solutions and, at the same time, supporting unidirectional acceleration and amplification of states is reminiscent of the skin effect~\cite{PhysRevLett.121.086803,PhysRevLett.77.570,PhysRevB.56.8651}. By further pushing the analogy with non-Hermitian linear systems, we expect that, in regime (III), stationary solitons will be present exclusively at one of the edges of the lattice, just as a linear system with a skin effect has eigenstates only at one of its edges. Indeed, when we adopt open boundary conditions (OBCs) and look for MF solutions living next to the right boundary, we obtain states that converge to machine precision and are linearly stable. The dynamical simulation in Fig.~\ref{DynamicSimulations}(c) confirms that such configurations remain static over long periods, and the plot in Fig.~\ref{DynamicSimulations}(f) shows that their norm is constant---not by necessity out of the equations of motion \eqref{OurModel}, but as a property of the solution. At the same time, there are no converging static solitons near the left boundary.

\emph{Discussion.} 
\begin{figure}
    \centering
    \includegraphics[width=\columnwidth]{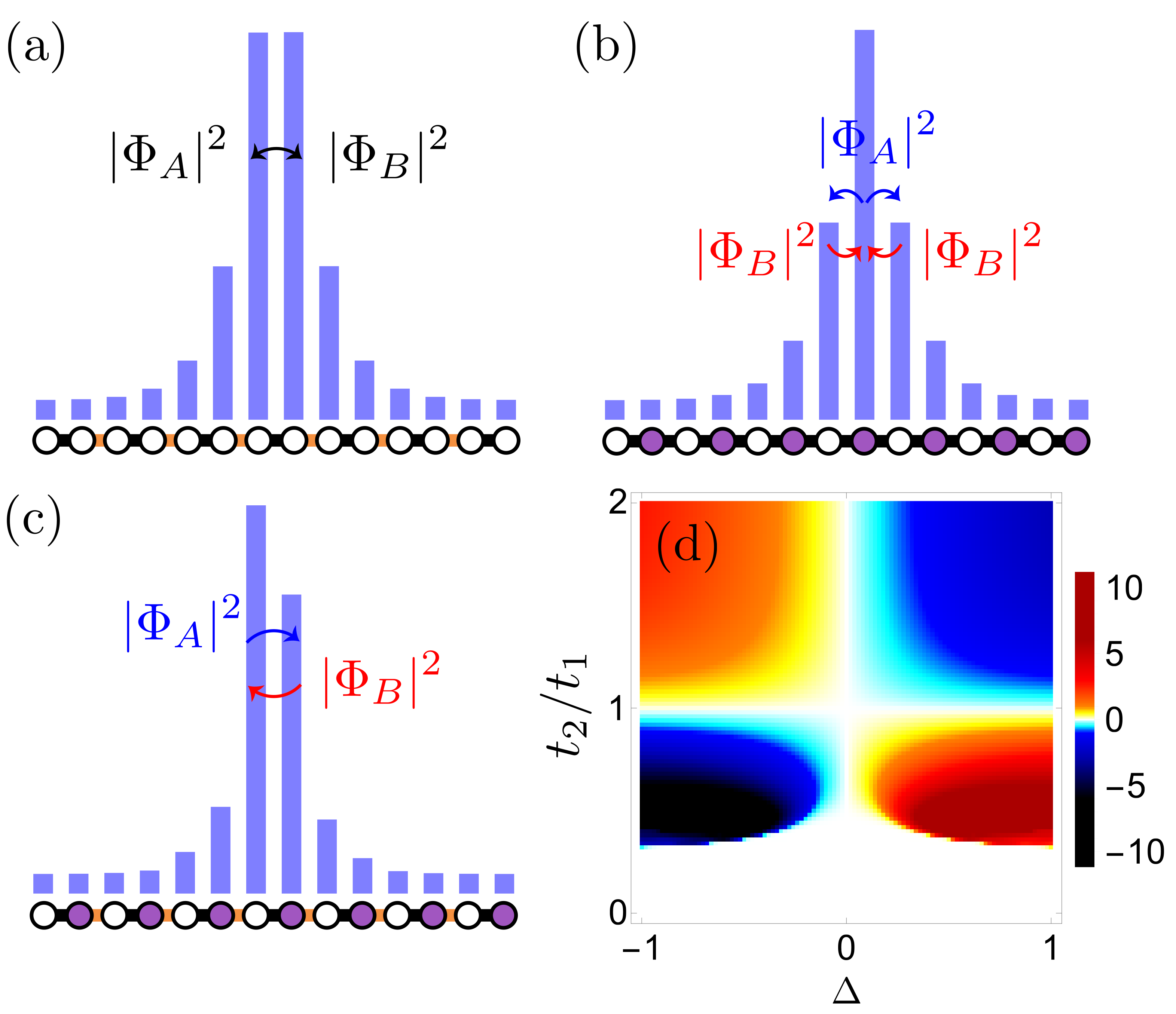}
    \caption{Parity breaking and nonreciprocity in our model, Eq.~\ref{OurModel}. (a)-(c) Circles and horizontal lines represent sites and linear couplings in the lattice. Blue vertical bars indicate the field amplitudes $|\Phi_{R,\alpha}|$ of representative solitons. (a) $t_1 \neq t_2$ (black and orange lines) and $\Delta=0$ (white circles); (b) $t_1 = t_2$ (black lines) and $\Delta \neq 0$ (white and purple circles); 
    (c) $t_1 \neq t_2$ and $\Delta \neq 0$. Only (c) breaks all parity symmetries, generating solitons susceptible to nonreciprocal dynamics. (d) Displacement of the center of mass (measured in unit cells) of initially static solitons corresponding to a time evolution from $t=0$ to $t=100$ as a function of the dimerization $t_2/t_1$ and the staggered energy $\Delta$. Colors represent the displacement for an initial configuration with $|\phi|^2=0.64$. Simulations are done with the RK4 method and a time step $\delta t=0.02$.}
    \label{LatticeDrawing}
\end{figure}
Setting $t_1=t_2$ and $\Delta=0$ in our system recovers the Ablowitz-Ladik (AL) equation, $\im\ddt\Phi_{R} = -(\epsilon+|\Phi_R|^2)(\Phi_{R+1}+\Phi_{R-1}) \label{AL}$, where $R \in \mathbb{Z}$ and $\epsilon \in \mathbb{R}$, which is known to support stable solitons~\cite{10.1063/1.522558,10.1063/1.523009}. Studies have found edge solitons and nonlinear skin modes in 1D lattices by introducing unequal parameters to the left- and right-nonlinear hopping terms of the AL equation~\cite{PhysRevB.104.L020303,PhysRevB.105.125421,Yuce_2021}, or employing nonreciprocal linear hoppings and an on-site Kerr nonlinearity~\cite{manda2024insensitiveedgesolitonsnonhermitian,PhysRevB.109.094308}. Another recent work~\cite{veenstra2024non} generated nonreciprocal solitons in a mechanical metamaterial using on-site nonlinearities and a nonreciprocal linear driving mechanism, similar to the Hatano-Nelson model~\cite{PhysRevLett.77.570,PhysRevB.56.8651}. In all cases, reciprocity is explicitly broken.

In contrast, our work shows the emergence of topological nonreciprocity from nonlinearity in the absence of parity. This nonreciprocity depends on the soliton's power and field profile. Since the nonlinear contribution of the AL equation for hopping from a site $R$ to $R+1$ is proportional to $|\Phi_R|^2$ whereas that for going in the opposite direction is proportional to $|\Phi_{R+1}|^2$ (Fig.~\ref{LatticeDrawing}), only the MF Hamiltonian of a non-uniformly distributed configuration of the field $\Phi$---as in a soliton---becomes non-Hermitian. In the original AL model, parity symmetry hinders the manifestation of nonreciprocity; all solitons are inversion symmetric about their center of mass~\cite{Discrete}, leading to nonlinear hopping amplitudes that cannot imbalance their motion in a particular direction. Hence, these solitons are always reciprocal. Figure \ref{LatticeDrawing} illustrates that setting $t_1 \neq t_2$ breaks the parity symmetry of the lattice about sites $A$ or $B$, and $\Delta \neq 0$ does the same about the center of the unit cell. The simultaneous dimerization and staggered on-site energies ensure that no solution with well-defined energy will be inversion symmetric. Then, the resulting unbalanced distribution of the field may accelerate the soliton as a whole. Figure \ref{LatticeDrawing}(d) shows how choosing $t_1/t_2$ and $\Delta$ determines the self-induced unidirectional propagation of initially static solitons. The lower white regions in Fig.~\ref{LatticeDrawing}(d) represent the points of parameter space where there is no self-acceleration as the solitons belong to regimes (I) and (II). In the Supplemental Material, we propose an experimental implementation of the AL equation and its parity-broken version for the observation of these unidirectional solitons.

Regardless of the particular details of our model, there are two critical elements for generating nonreciprocal dynamics in nonlinear periodic lattice models: (i) non-Hermiticity, which automatically breaks TRS or $\mathrm{TRS}^{\dagger}$, and (ii) parity symmetry breaking. Thus, more generally, any lattice that contains the same sort of nonlinear coupling and breaks parity symmetry could also exhibit nonreciprocal solitons. In the Supplemental Material, we show this to be the case for the Salerno model~\cite{PhysRevA.46.6856}, which contains AL and Gross-Pitaevskii contributions to the nonlinear Schr\"odinger equation. In addition, since the nonreciprocal solitons appear in regime (III) where $|\phi|^2$ is small, it is natural to ask if such states endure as the field amplitude decreases enough to enter the quantum realm. The quantized version of the AL model~\cite{Chowdhury_1993} is an example of a $q$-boson model~\cite{10.1063/1.522937,Biedenharn_1989,Macfarlane_1989} and the effects of parity breaking on this kind of strongly interacting bosonic system are still unknown.

\emph{Acknowledgments.} We thank Vir Bulchandani, Nicolas Regnault, Shinsei Ryu, Frank Schindler, and Ajit Srivastava for discussions on related topics. W.A.B. is thankful for the support of the startup funds from Emory University and by Grant No. NSF PHY-1748958 to the Kavli Institute for Theoretical Physics (KITP) during the program ``A Quantum Universe in a Crystal: Symmetry and Topology across the Correlation Spectrum".

\emph{Data availability.} All data is available upon reasonable request.

\bibliography{main}

\end{document}


\title{
Solitons with Self-induced Topological Nonreciprocity \\ Supplemental Material  
}

\author{Pedro Fittipaldi de Castro and Wladimir A. Benalcazar}
\affiliation
{Department  of  Physics,  Emory  University, Atlanta, GA 30322, USA}

\date{\today}

\maketitle

In this Supplemental Material, we review the Hamiltonian formulation of the AL equation in section I and comment on the fate of symmetries and constants of motion as we break parity in the linear terms of the AL model. In section II, we discuss in detail our method for numerically obtaining solitons in the parity-broken AL model. In section III, we discuss the linear stability analysis of those solutions and, in section IV, we define a way to calculate the winding number of the mean field and stability spectra. Section V shows the behavior of the point gaps in the MF a and stability spectra as we transition from regime (II) to (III). In section VI, we show a dynamical simulation of a nonreciprocal soliton in the parity-broken Salerno model, which has on-site nonlinearities as well as AL terms.
In section VII, we compare the spectra and local density of states in the parity-broken AL model with a linear ``local Hatano-Nelson mode'', showing that the solitons act as ''nonreciprocal defects'' and are responsible for a nonlinear skin effect. Section VIII discusses the long-time evolution of nonreciprocal solitons and section IX outlines a possible experimental realization of the parity-broken AL model in topolectric circuits.

\section{Hamiltonian formalism of the AL equation}

We can establish a Hamiltonian formulation of the AL equation,
\begin{equation}
    \im \ddt \Phi_R = -(\epsilon+|\Phi_R|^2)(\Phi_{R+1}+\Phi_{R-1}), \label{equationAL}
\end{equation}
by taking $\Phi_R$ and $\Phi^{*}_R$ as canonically conjugate variables and introducing the weighted Poisson bracket
\begin{equation}
    \poissonbracket{f}{g}_{AL} = \im \sum_{R=0}^{N-1}(\epsilon+|\Phi|_R^2)\Big(\pdv{f}{\Phi_R^*}\pdv{g}{\Phi_R}-\pdv{f}{\Phi_R}\pdv{g}{\Phi_{R}^{*}}\Big). \label{Poisson}
\end{equation}
The Hamiltonian of the AL system is a constant of motion and it is given by
\begin{equation}
    H_{AL} = -\sum_{R=0}^{N-1}\left(\Phi_{R}\Phi_{R+1}^* - \Phi_{R}^*\Phi_{R+1}\right)
\end{equation}
Then, one can check that $\ddt \Phi_R = \{\Phi_R,H_{AL}\}_{AL}$ is equivalent to the equation of motion \eqref{equationAL}. It is also easy to show \cite{Cuevas-Maraver_2019} that the norm-like quantity
\begin{equation}
     P = \sum_R \ln\left(1+\frac{1}{\epsilon}|\Phi_R|^2\right) \label{ConservedQuantity}
\end{equation}
is also conserved.

\subsection{Effects of breaking parity}

If we explicitly break parity in the linear terms of the AL equation \eqref{equationAL}, we obtain the following equations of motion:
\begin{equation}
    \im\ddt \Phi_a = -\sum_{b}(h_{ab}+|\Phi_a|^2 f_{ab})\Phi_{b}
    \label{OurModel}
\end{equation}
where $h$ and $f$ are the tight-binding matrices,
\begin{align}
    h_{(R,\alpha)(R',\beta)} &= \left(t_1\sigma^x_{\alpha\beta}+\Delta\sigma^z_{\alpha\beta}\right)\delta_{R,R'} +t_2\left(\sigma^{+}_{\alpha\beta}\delta_{R-1,R'}+\sigma^{-}_{\alpha\beta}\delta_{R+1,R'}\right) \nonumber \\
    f_{(R,\alpha)(R',\beta)} &= (\sigma^{x}_{\alpha\beta}\delta_{R,R'}+\sigma^{+}_{\alpha\beta}\delta_{R-1,R'}+\sigma^{-}_{\alpha\beta}\delta_{R+1,R'}). \label{matrices}
\end{align}

Now, the weighted Poisson bracket defined in \eqref{Poisson} is not applicable anymore, and it is not even clear whether there is an adequate Hamiltonian formulation of the parity-broken AL model. In addition, there is no clear generalization of the quantity $P$ in \eqref{ConservedQuantity}. Therefore, since there is no notion of a conserved norm-like function in the parity-broken AL model, we adopt $|\Phi|^2$ as the definition of the norm of the configuration $\Phi$. The evolution of the squared amplitude has a natural physical interpretation as the field intensity, and $|\Phi|^2$ is also a direct measure of the strength of the nonlinear terms in the equations of motion \eqref{OurModel}.

\section{Self-consistent solutions}

The key to exposing nonreciprocal behavior in our model is to investigate the fate of static bulk solitons as we vary the parity-breaking parameters $t_2/t_1$ and $\Delta$. We can easily find these solutions by changing variables to the rotating frame $\Phi_a(t)=e^{-\im \omega t}\phi_a(t)$, where $\omega \in \mathbb{R}$ is some real frequency to be determined~\cite{schindler2023nonlinear}. In the rotating frame, the equations of motion \eqref{OurModel} take the form
\begin{equation}
    \im\ddt \phi_a = -\sum_{b}\left(h_{ab}-\omega\delta_{ab}+|\Phi_a|^2 f_{ab}\right)\Phi_{b} \label{RotatingFrame}
\end{equation}
Then, it is clear that static solitons are stationary in the rotating frame, $\ddt \phi_a=0$~\cite{FLACH1998181}, yielding the time-independent expression
\begin{align}
    \omega \phi_a &= -\sum_{b}(h_{ab}+|\phi_a|^2 f_{ab})\phi_{b} \equiv \sum_b H_{ab}[\phi]\phi_{b}, \label{SelfConsistent}
\end{align}
which we treat as a linearized equation for the ``eigenstate'' $\phi$ and ``eigenvalue'' $\omega$, and which we self-consistently solve by choosing a localized initial guess $\phi^{(0)}$, diagonalizing the effective Hamiltonian $H[\phi^{(0)}]$, and picking the resulting eigenvector $\phi^{(1)}$ having the highest overlap with the initial guess. Then, we iterate the process, obtaining a sequence $\{\phi^{(i)}\}$ that we accept to converge once we reach an iteration $n$ such that $|\phi^{(n)}-\phi^{(n-1)}|<\delta \sim 10^{-15}$ (machine precision).
The outcome of the process is a soliton $\phi$ with energy (frequency) $\omega$ and a self-consistent, mean-field Hamiltonian $H[\phi]$, which, in addition to $\phi$, has a multitude of extended eigenvectors $\psi$ with energies $\varepsilon$.

\section{Stability analysis}

To build the stability matrix $M$ associated with a static soliton $\phi$ and solve for its eigenvalues, we write the equations of motion in the rotating frame \eqref{RotatingFrame} explicitly:
\begin{align}
    \im\ddt\phi_{R,A} = &-\omega\phi_{R,A} -\Delta \phi_{R,A} - t_1\phi_{R,B} - t_2\phi_{R-1,B} - |\phi_{R,A}|^2(\phi_{R,B}+\phi_{R-1,B}) \nonumber \\
    \im\ddt\phi_{R,B} = &-\omega\phi_{R,B} +\Delta \phi_{R,B} - t_1\phi_{R,A} - t_2\phi_{R+1,A} - |\phi_{R,B}|^2(\phi_{R,A}+\phi_{R+1,A}). \label{ExplicitEquations}
\end{align}
Now, we add a time-dependent fluctuation $\delta\phi(t)$ to the solution, yielding a perturbed soliton
\begin{equation}
   \Tilde{\phi}_{R,\alpha}=\phi_{R,\alpha}+\delta\phi_{R,\alpha}(t). \label{Perturbed}
\end{equation}
Then, we substitute \eqref{Perturbed} back into \eqref{ExplicitEquations} and collect the terms in the resulting expression that are linear in $\delta\phi$, yielding the linearized differential equation for the fluctuations
\begin{align}
    \im\ddt\delta\phi_{R,A} = &-(\omega+\Delta)\delta\phi_{R,A} - t_1\delta\phi_{R,B} - t_2\delta\phi_{R-1,B} - |\phi_{R,A}|^2(\delta\phi_{R,B}+\delta\phi_{R-1,B}) \nonumber \\
    & \ \ - (\phi_{R,B}+\phi_{R-1,B})(\phi^*_{R,A}\delta_{R,A}+\phi_{R,A}\delta\phi^*_{R,A})\nonumber \\
    \im\ddt\delta\phi_{R,B} = &(-\omega+\Delta)\delta\phi_{R,B} - t_1\delta\phi_{R,A} - t_2\delta\phi_{R+1,A} - |\phi_{R,B}|^2(\delta\phi_{R,A}+\delta\phi_{R+1,A}) \nonumber \\
    & \ \ - (\phi_{R,A}+\phi_{R+1,A})(\phi^*_{R,A}\delta_{R,A}+\phi_{R,A}\delta\phi^*_{R,A}). \label{Linearized}
\end{align}
We can write the complex fluctuations, without loss of generality, as $\delta\phi_{R,\alpha}(t) = v_{R,\alpha}e^{\Lambda t}+w^*_{R,\alpha}e^{\Lambda^*t}$, where $v_{R,\alpha}$ and $w_{R,\alpha}$ are time-independent complex numbers. By substituting this expression in \eqref{Linearized} and collecting the terms proportional to $e^{\Lambda t}$ and $e^{\Lambda^*t}$, we get a set of $4L$ linear equations:
\begin{align}
    \im \Lambda v_{R,A} &= -(\omega+\Delta)v_{R,A} - t_1 v_{R,B} - t_2 v_{R-1,B} - |\phi_{R,A}|^2(v_{R,B}+v_{R-1,B})-(\phi_{R,B}+\phi_{R-1,B})(\phi^*_{R,A}v_{R,A}+\phi_{R,A}w_{R,A}) \nonumber \\
    \im \Lambda v_{R,B} &= (-\omega+\Delta)v_{R,B} - t_1 v_{R,A} - t_2 v_{R+1,A} - |\phi_{R,B}|^2(v_{R,A}+v_{R+1,A})-(\phi_{R,A}+\phi_{R+1,A})(\phi^*_{R,B}v_{R,B}+\phi_{R,B}w_{R,B}) \nonumber \\
    \im \Lambda w_{R,A} &= (\omega+\Delta)w_{R,A} + t_1 w_{R,B} + t_2 w_{R-1,B} + |\phi_{R,A}|^2(w_{R,B}+w_{R-1,B})+(\phi^*_{R,B}+\phi^*_{R-1,B})(\phi_{R,A}w_{R,A}+\phi^*_{R,A}v_{R,A}) \nonumber \\
    \im \Lambda w_{R,B} &= (\omega-\Delta)w_{R,B} + t_1 w_{R,A} + t_2 w_{R+1,A} + |\phi_{R,B}|^2(w_{R,A}+w_{R+1,A})+(\phi_{R,A}+\phi_{R+1,A})(\phi^*_{R,B}w_{R,B}+\phi_{R,B}v_{R,B})
\end{align}
which we can compactly express as
\begin{equation}
    \lambda 
    \begin{pmatrix}
        v \\
        w
    \end{pmatrix}
    =
    M
    \begin{pmatrix}
        v \\
        w
    \end{pmatrix}
    , \quad 
    M = \begin{pmatrix}
        \mathcal{A} & \mathcal{B} \\
        -\mathcal{B}^* & - \mathcal{A}^*
    \end{pmatrix},
\end{equation}
where $v=(v_{1,A},v_{1,B},\dots,v_{L,A},v_{L,B})$ and the same for $w$, $\lambda = i\Lambda$ and
\begin{align}
    \mathcal{A}_{(R,\alpha),(R',\beta)} &= H[\phi]_{(R,\alpha),(R',\beta)} - \left[ \omega + \phi_{R,\alpha}^*\sum_{R'',\gamma}f_{(R,\alpha),(R'',\gamma)}\phi_{(R'',\gamma)} \right]\delta_{R,R'}\delta_{\alpha,\beta}, \nonumber \\
    \mathcal{B}_{(R,\alpha),(R',\beta)} &= -\phi_{R,\alpha}\sum_{R'',\gamma}f_{(R,\alpha),(R'',\gamma)}\phi_{(R'',\gamma)}\delta_{R,R'}\delta_{\alpha,\beta} \label{ABmatrices}.
\end{align}

\section{Winding number}

The winding number $\nu$ of a closed curve in the plane relative to a given point is an integer counting the number of times that curve travels counterclockwise around that point. If two curves can be continuously deformed into one another without crossing the reference point, then their winding number is the same. In that sense, $\nu$ is a topological invariant. A generic formulation of the winding number goes as follows: let
\begin{align}
    &\gamma:[t_0,t_1] \to \mathbb{C}\nonumber \\
    &t \mapsto z(t) = r(t)e^{i\theta(t)}
\end{align}
be a continuously differentiable closed, parametrized curve on the complex plane where $r$ and $\theta$ are the polar coordinates. Now, consider the differential
\begin{equation}
    \frac{\diff z}{z} = \frac{\diff r}{r} + \im\diff \theta = \diff (\log r) + \im\diff\theta.
\end{equation}
Integrating the quantity above along the curve $\gamma$ from $t_0$ to $t_1$ gives
\begin{equation}
    \int_{t_0}^{t_1} \frac{\dot{z}(t)}{z(t)}\diff t = \int_{r(t_0)}^{r(t_1)} \diff(\log r) + \im \int_{t_0}^{t_1} \diff\theta = \log\left(\frac{r(t_1)}{r(t_0)}\right) + \im (\theta(t_1)-\theta(t_0)).
\end{equation}
But because $\gamma$ is a closed curve, $r(t_1)=r(t_0)$ and $\theta(t_1)=\theta(t_0)+2\pi\nu$ $(\nu \in \mathbb{Z})$, so we have
\begin{equation}
    \nu = \frac{1}{2\pi \im} \int_{t_0}^{t_1} \frac{\dot{z}(t)}{z(t)}\diff t = \frac{1}{2\pi \im} \oint_{\gamma} \frac{\diff z}{z},
\end{equation}
which counts how many times $\gamma$ winds around the origin of the complex plane. By shifting the origin of the polar coordinates $z(t) = z_0 + r(t)e^{\im\theta(t)}$, we can easily generalize the above expression to represent the winding number of $\gamma$ relative to a generic point $z_0 \in \mathbb{C}$:
\begin{equation}
    \nu(\gamma;z_0) = \frac{1}{2\pi\im}\int_{t_0}^{t_1}\frac{\dot{z}(t)dt}{z(t)-z_0} = \frac{1}{2\pi \im} \oint_{\gamma} \frac{\diff z}{z-z_0}, 
\end{equation}
which is nothing more than a particular case of Cauchy's integral formula.

One may generalize the idea of the winding number of a curve to the winding of a linear operator and, as a result, formulate a topological classification of any non-Hermitian Hamiltonian $H^{\dagger} \neq H$, provided that there is a parameter $t\in \mathbb{R}$ such that $\det H$ is periodic in $t$, that is, $\det H(t+T) = \det H(t)$~\cite{PhysRevX.9.041015}\cite{PhysRevB.105.165137}. This way, the eigenvalues $E_i(t)$ will each have a contribution $\nu_i$ (in general, not an integer) to the winding of the spectrum of $H$ as $t$ goes from $0$ to $T$. The total winding is simply the sum of all the individual $\nu_i$ and must add up to an integer:
\begin{align}
    \nu &\equiv \frac{1}{2\pi \im} \oint_{0}^{T} \diff t  \ \partial_{t} \ln \det H(t) \nonumber \\
    &= \frac{1}{2\pi \im} \oint_{0}^{T} \diff t  \ \partial_{t} \Tr \ln H(t) \nonumber \\
    &= \frac{1}{2\pi \im} \oint_{0}^{T} \diff t  \Tr[H^{-1}(t)\partial_{t}H(t)] \nonumber \\
    &= \frac{1}{2\pi \im} \sum_{i} \oint_{0}^{T} \diff t \frac{\partial_t E_i(t)}{E_i(t)} = \sum_{i} \nu_i,
\end{align}
where we have assumed that $\det H(t) \neq 0 \ \forall \ t \in [0,T]$. Notice that this formulation does not rely on translation symmetry and the index $i$ runs over all eigenvalues $E_i$ of $H$, so it is not a band index. One can immediately generalize the above result to an arbitrary reference energy $E \neq 0$ by replacing $H(t)$ with $H(t) - E_0$, so that the constraint becomes $\det[H(t)-E_0] \neq 0 \  \forall t \ \in [0,T]$. 

\subsection{Application to the parity-broken AL model}

One may define the winding number of a non-Hermitian Hamiltonian even in the absence of translational symmetry by coupling the system to a fictitious gauge field (which is possible in our system due to its U(1) symmetry)~\cite{PhysRevX.9.041015}\cite{PhysRevB.105.165137}. For the sake of simplicity, let us swap our $(R,\alpha)$ notation for a single integer index as
\begin{equation}
    \begin{pmatrix}
        \phi_{1} \\
        \phi_{2} \\
        \vdots \\
        \phi_{N-1} \\
        \phi_{N}
    \end{pmatrix}
    =
    \begin{pmatrix}
        \phi_{1,A} \\
        \phi_{1,A} \\
        \vdots \\
        \phi_{L,A} \\
        \phi_{L,B}
    \end{pmatrix}
\end{equation}
with $N=2L$, such that the $A$ and $B$ sublattices map to the odd and even positive integers, respectively. Now, let us introduce a complex phase factor in the elements of the MF Hamiltonian as
\begin{equation}
    H_{mn} \mapsto H_{mn}(\theta) = e^{\im\theta([m]-[n])/N}H_{mn},
\end{equation}
where $m,n = 1,\dots,N$ and $[m]=m \mod{N}$, so that the total flux across the lattice is $\theta$. From this definition, the gauge-invariant quantity $\det H(\theta)$ is periodic in $\theta$ with period $2\pi$ and we can define the winding number relative to a reference energy $E_0 \in \mathbb{C}$ as
\begin{equation}
    \nu(E) \equiv \frac{1}{2\pi\im} \int_{0}^{2\pi} \diff \theta \ \partial_{\theta} \ln \det \left[H(\theta)-E_0\right] \;\; \in \mathbb{Z}. \label{winding}
\end{equation}

The stability matrix is a $2N \times 2N$ matrix formed by four $N \times N$ blocks that inherit the periodicity of the original lattice with $N$ sites. Therefore, we consider the following transformation:

\begin{equation}
    M_{mn} \mapsto M_{mn}(\theta) = e^{\im\theta([m]-[n])/N}M_{mn},
\end{equation}
and the winding number of the stability spectrum relative to a point $\lambda_0$ in the complex plane is
\begin{equation}
    \nu(\lambda) \equiv \frac{1}{2\pi\im} \int_{0}^{2\pi} \diff \theta \ \partial_{\theta} \ln \det \left[M(\theta)-\lambda_0\right] \;\; \in \mathbb{Z}. \label{windingstab}
\end{equation}

\begin{figure}
    \centering
    \includegraphics[width=10cm]{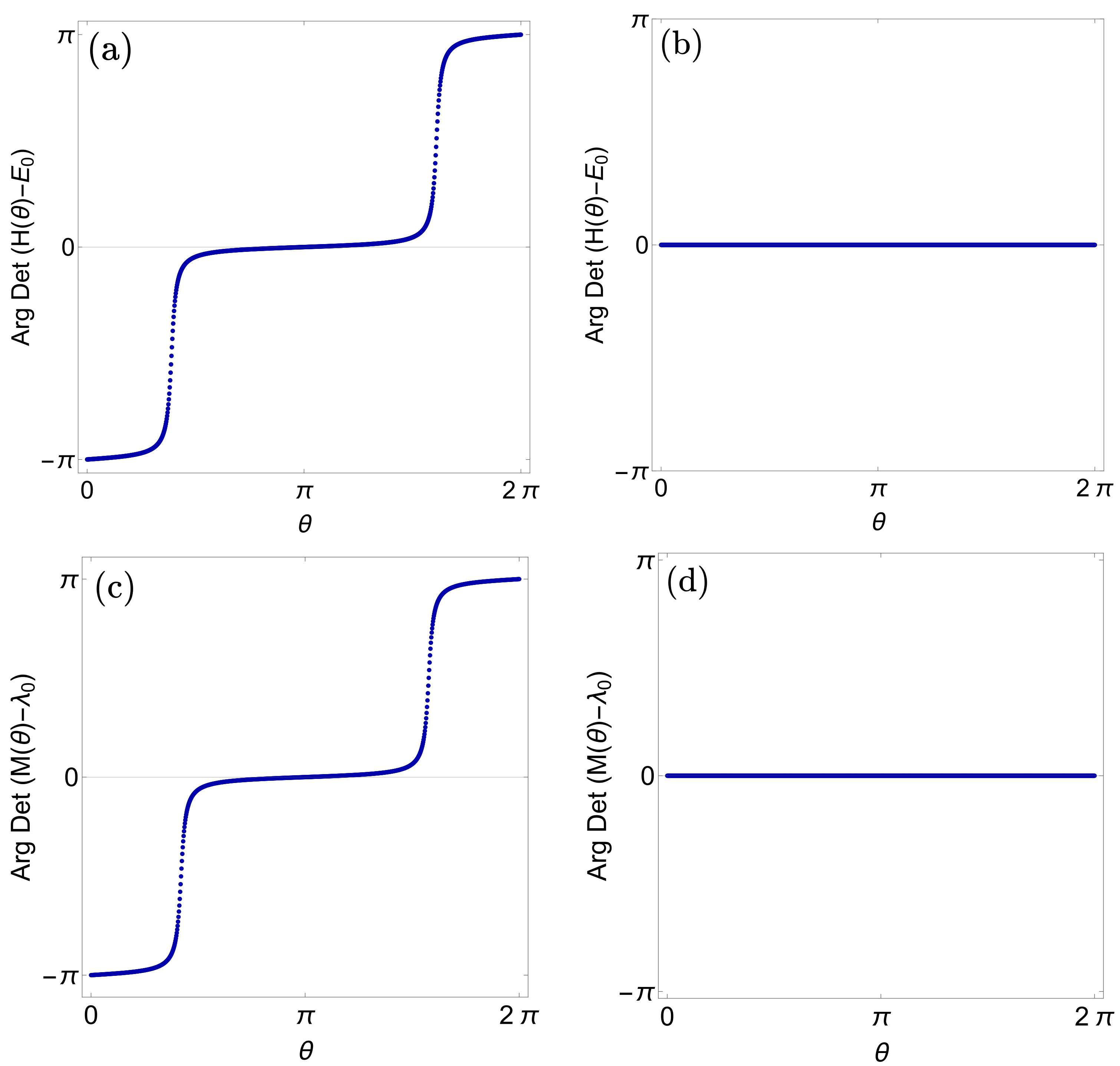}
    \caption{Winding of the MF and stability spectra. (a) Advance of the argument of $\ln \det (H(\theta-E_0))$ for a reference energy $E_0=1.7$ inside the point gap of the MF spectrum of a soliton with power $|\phi|^2=1$. (b) Same as (a) but for a reference energy $E=0$ outside the point gap. (c) Advance of the argument of $\ln \det (M(\theta-\lambda_0))$ for a reference stability value $\lambda_{0}=3.8$ inside the point gap of the stability matrix of a soliton with power $|\phi|^2=1$. Same as (c) but for a reference value $\lambda_0=0$ outside the point gap.}
    \label{Winding}
\end{figure}

The numerical evaluations of \eqref{winding}  and \eqref{windingstab} give $\nu=\pm 1$ as long as the point gap is open about the references $E_0$ and $\lambda_0$. Figs. \eqref{Winding}(a) and (c) show how the arguments of $\ln \det \left[H(\theta)-E_0\right]$ and $\ln \det \left[M(\theta)-\lambda_0\right]$ advance smoothly from $-\pi$ to $\pi$ as $\theta$ goes from 0 to $2\pi$ when the point gap is open. Figs. (a) and (c) show how the winding vanishes outside the point gap.

\section{Mean-field energy and stability bands}

\begin{figure}
    \centering
    \includegraphics[width=11.5cm]{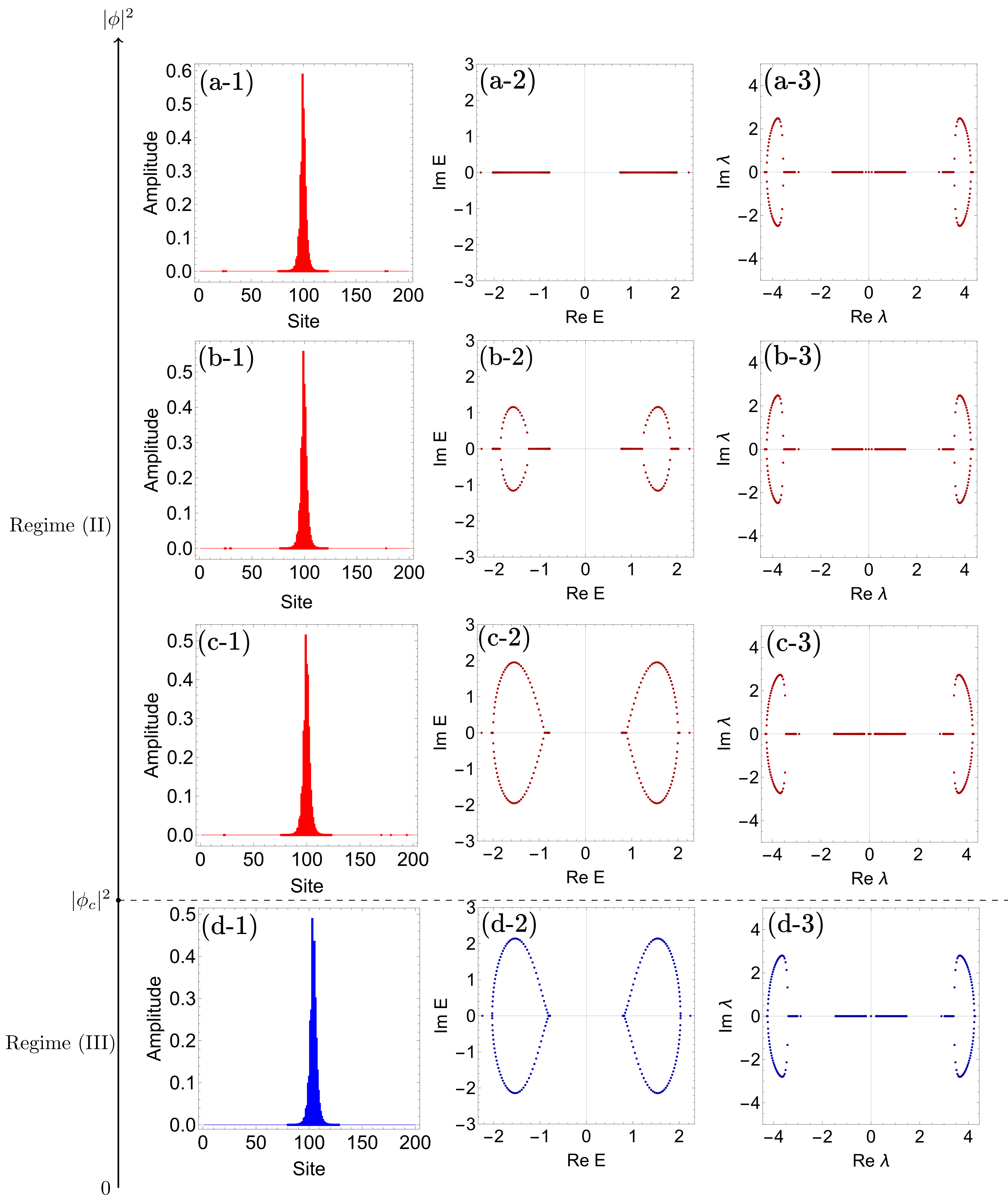}
    \caption{Transition between regimes (II) and (III). (a-1,2,3) Soliton profile, MF spectrum, and stability spectrum at $|\phi|^2=1.05$, respectively, with imaginary parts scales by a factor of $10^{4}$. (b-1,b-2,b-3) Same as (a-1,2,3) but for $|\phi|^2=1$. (c-1,c-2,c-3) Same as (a-1,2,3) but for $|\phi|^2=0.94$. (d-1,d-2,d-3) Same as (a-1,2,3) but for $|\phi|^2=0.93$. Parameters used in the numerical calculations: $t_1=1.35$, $t_2=0.65$, $\Delta=0.35$, and $L=100$ unit cells.}
    \label{MFenergies}
\end{figure}

In the main text, we claim that the point gaps inside the MF energy bands reach the outermost edges $E_{\pm}$ at the transition point $|\phi|^2=0.94$. In Fig. \ref{MFenergies}, we show that the point gaps within regime (II) grow as we lower the soliton power, eventually reaching the band edges. At regime (III) $|\phi|^2<0.94$, the self-consistent method for finding stationary solutions does not converge. However, for the sake of completeness, we show the resulting state and spectra after $\sim 100$ iterations for $|\phi|^2=0.93$ in subfigures \ref{MFenergies}(c-1,2,3). There, we observe that the soliton profile is slightly displaced to the right when compared to the plots above it, showing that the lack of convergence is due to the tendency of solitons in that regime to move in that direction. If we increase the number of iterations, the resulting profile will translate even further to the right relative to the initial ansatz (placed at the center).

\begin{figure}[h]
    \centering
    \includegraphics[width=0.7\linewidth]{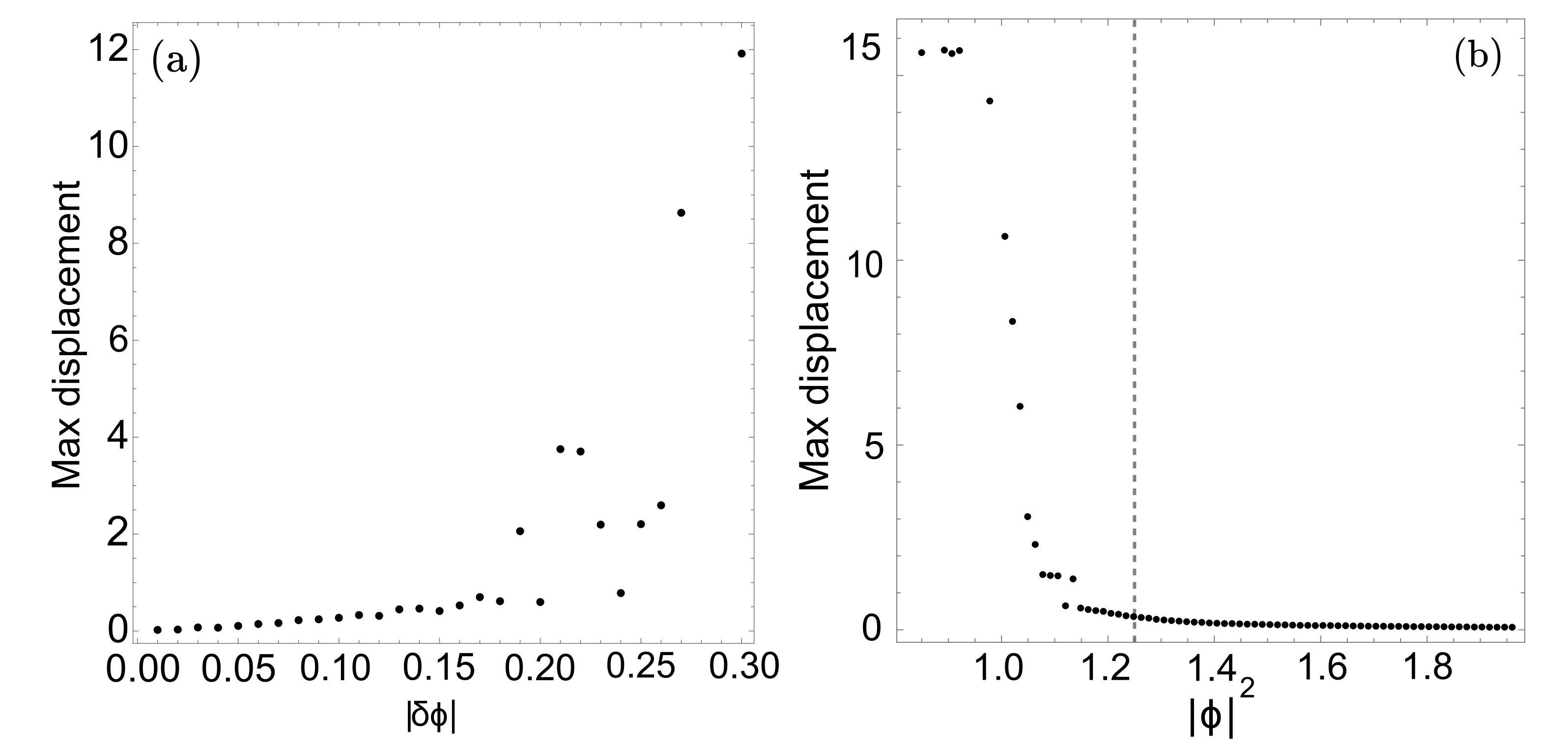}
    \caption{Validity of perturbative stability analysis (a) Maximum displacement (measured in unit cells) of an initially static soliton with $|\phi|^2=1.27$ slightly above the transition between regimes (I) and (II) as a function of the norm $|\delta\phi|$ of the random initial perturbation. (b) Maximum displacement (measured in unit cells) of initially static solitons in the presence of random perturbations as a function of their field intensity after a time evolution from $t=0$ to $t=2000$ ($\sim 1000$ periods of phase oscillation). Each black dot is the average result 20 simulations with the same initial norm but independent random perturbations. The dashed gray line indicates the transition value between regimes (I) and (II) in the first order in perturbation theory. We have done those simulations in a lattice with 20 unit cells under OBC, which sets the largest maximal displacement of the center of mass of the soliton to 15 unit cells.}
    \label{DisplacementA}
\end{figure}

Regarding the transition between regimes (I) and (II), since our analysis is first-order in perturbation theory, higher-order contributions of a perturbation could in principle affect the transition points of soliton behavior. To test the transition between regimes (I) and (II), we performed dynamic simulations of initially static solitons with additional random perturbations. The results are shown in Figs. \ref{DisplacementA}a,b. Our prediction for the phase transition point between regimes (I) and (II) (dotted vertical line at $|\phi|^2=1.25$) reliably separates these two regimes (see Fig. \ref{DisplacementA}b), suggesting that higher-order terms in the perturbations do not (appreciably) modify our results for small perturbations (see Fig. \ref{DisplacementA}a).

\section{Nonreciprocal solitons in the Salerno model}

Beyond the minimal model discussed in the main text, which only contains AL nonlinearities, we observe that the solitons' topological nonreciprocity persists upon the addition of on-site nonlinear contributions to the equations of motion \eqref{ExplicitEquations}:

\begin{align}
    \im\ddt\phi_{R,A} = &-\omega\phi_{R,A} -\Delta \phi_{R,A} - t_1\phi_{R,B} - t_2\phi_{R-1,B} - g_0|\phi_{R,A}|^2\phi_{R,A} - g_1|\phi_{R,A}|^2(\phi_{R,B}+\phi_{R-1,B}) \nonumber \\
    \im\ddt\phi_{R,B} = &-\omega\phi_{R,B} +\Delta \phi_{R,B} - t_1\phi_{R,A} - t_2\phi_{R+1,A}  - g_0 |\phi_{R,B}|^2\phi_{R,B}- g_1|\phi_{R,B}|^2(\phi_{R,A}+\phi_{R+1,A}), \label{SalernoModel}
\end{align}
where $g_0$ and $g_1$ control the strengths of the on-site and AL nonlinearities, respectively. The resulting equations \eqref{SalernoModel} describe a parity-broken version of the Salerno model~\cite{PhysRevA.46.6856}, which interpolates between the Gross-Pitaevskii and AL equations. 

\begin{figure}
    \centering
    \includegraphics[width=14cm]{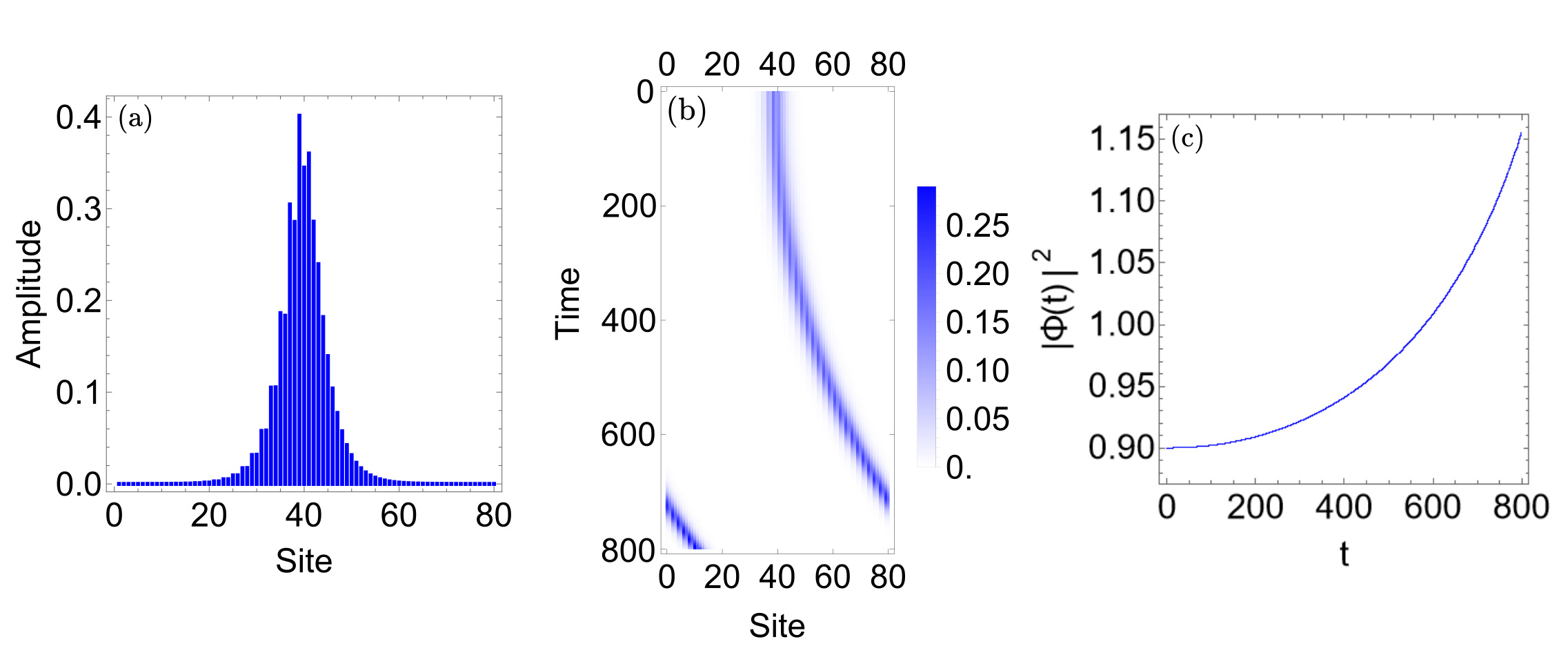}
    \caption{Nonreciprocal soliton in the Salerno model. (a) Field profiles $\phi_R$ at $t=0$. (b ) Evolution of the square amplitude per site from $t=0$ to $t=800$ under PBC. (c) Evolution of the soliton's power from $t=0$ to $t=800$.
    Simulation made with the 4th order Runge-Kutta algorithm with a time step $\Delta t = 0.01$ and lattice parameters $t_1=1.5$, $t_2=0.5$, $\Delta=0.3$, on-site nonlinearity $g_0=0.6$ and AL nonlinearity $g_1=0.2$. System size $L=40$ unit cells ($80$ sites).}
    \label{Salerno}
\end{figure}

Fig. \ref{Salerno} displays the evolution of a soliton according to eq. \eqref{SalernoModel}, where we observe acceleration and amplification even when $g_1/g_0 \sim 0.1$. This shows the robustness of topological nonreciprocity upon the addition of on-site nonlinearities to the equations of motion.

\begin{figure}
    \centering
    \includegraphics[width=0.8\linewidth]{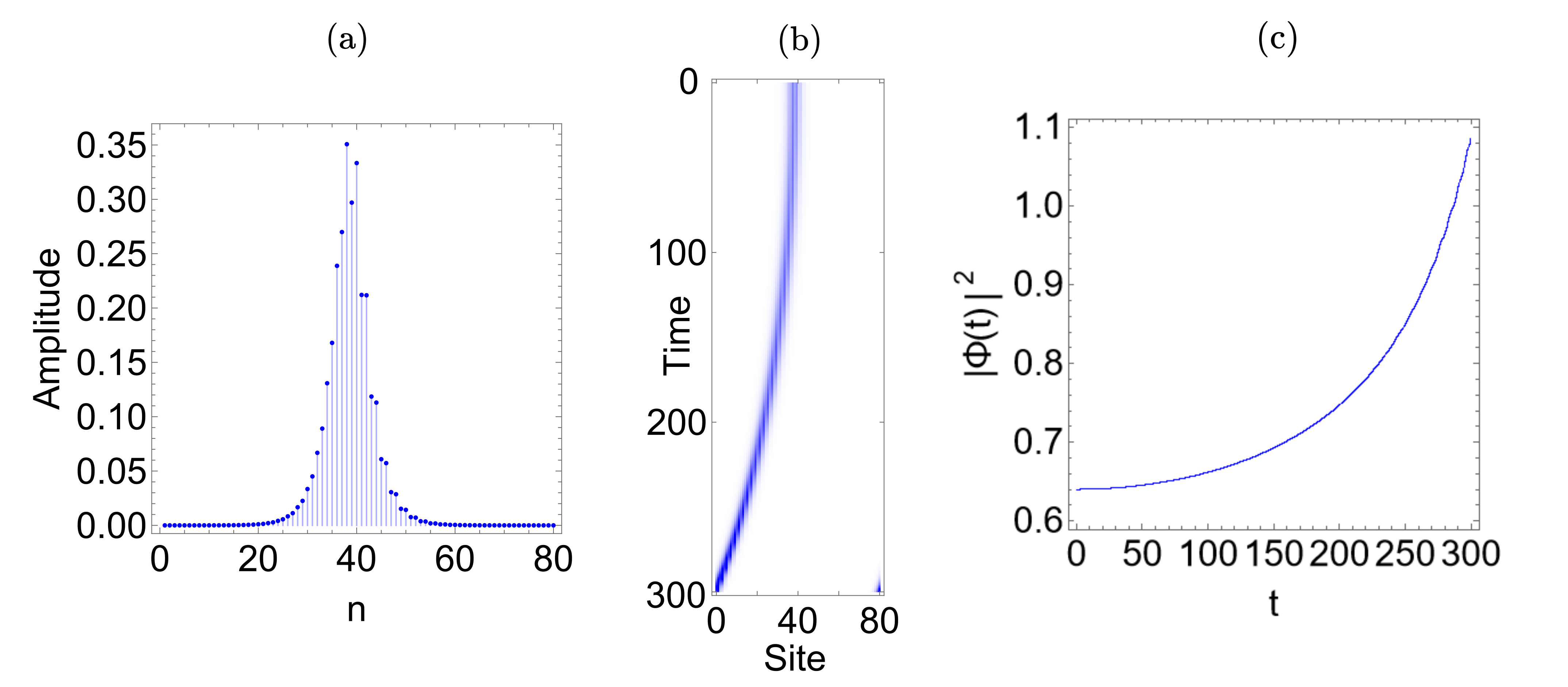}
    \caption{Soliton with MF and stability winding numbers $\nu=-1$. (a) Initial soliton profile. (b) Intensity per site as a cuntion of time. (c) Total intensity as a function of time.}
    \label{Minus1Soliton}
\end{figure}

For the sake of completeness, in figure Fig. \ref{Minus1Soliton} we show the time evolution of a nonreciprocal soliton in the parity-broken AL model whose winding number is $\nu=-1$, showing that it accelerates and amplifies in the opposite direction as the solitons with $\nu=+1$.

\section{Local Hatano-Nelson model and connection to the skin effect}

In the main body of the paper, we argue that the regime exhibiting unidirectional acceleration of solitons in the parity-broken AL model is a manifestation of a nonlinear skin effect generated by the nonreciprocal effective hopping amplitudes localized in the region of support of the soliton. We justify such an interpretation by comparing the results of our nonlinear model to a linear system having a ``non-Hermitian defect'', \emph{i.e.}, a localized region $A$ of the lattice where the left and right linear hopping strengths are unequal. We call such a system a local Hatano-Nelson model, whose Hamiltonian is given by
\begin{align}
    H_{LHN} &= -t\sum_{R \in A^C} (\ket{R}\bra{R+1} + \ket{R+1}\bra{R})  - \sum_{R \in A} (t_l\ket{R}\bra{R+1}+t_r\ket{R+1}\bra{R}),
\end{align}
where $t_l$ and $t_r$ are the left and right hopping strengths on the nonreciprocal defect $A$ and $t$ is the hopping parameter of the Hermitian part of the lattice $A^C$.

\begin{figure}
    \centering
    \includegraphics[width=.6\textwidth]{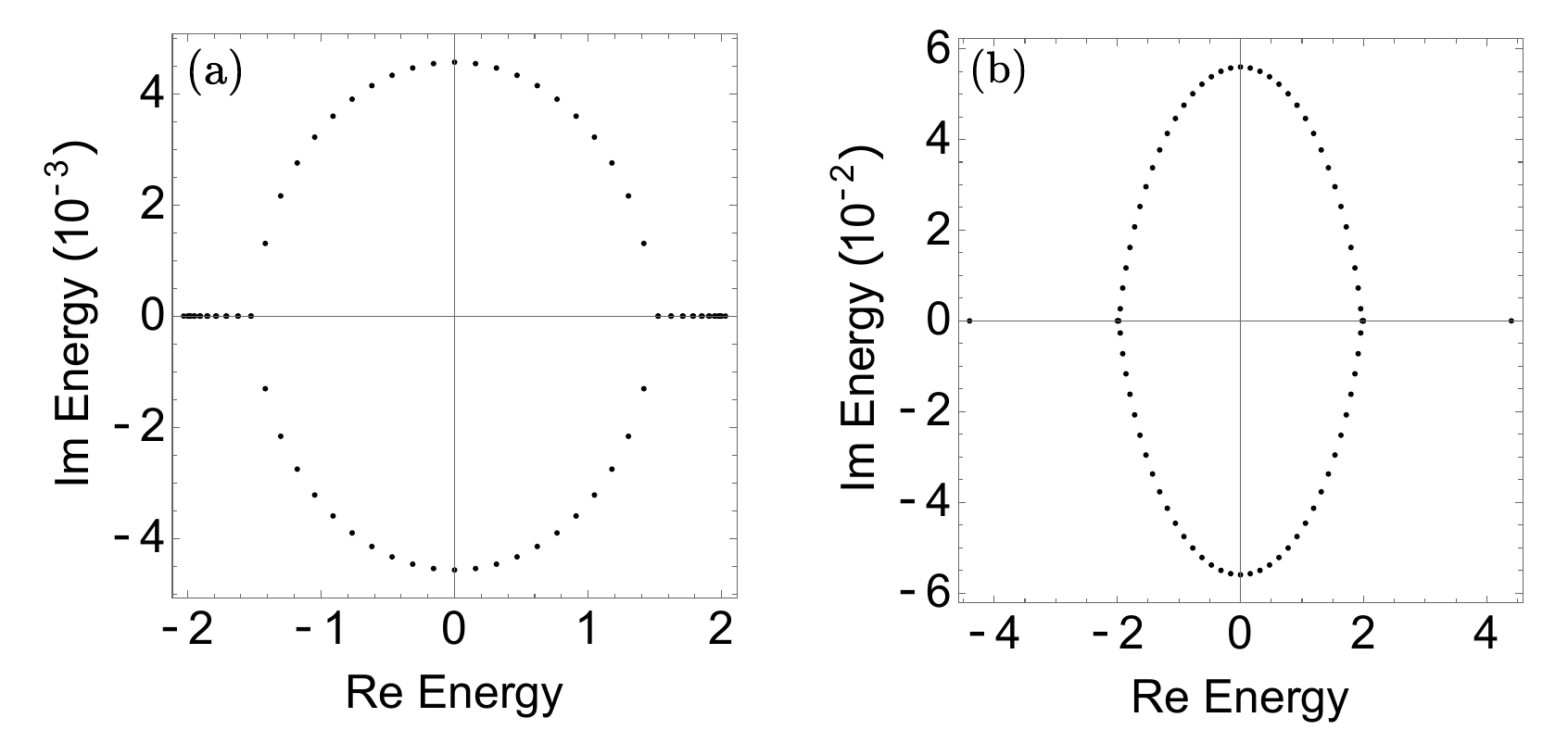}
    \caption{Energy spectrum of the LHN model, where $A=2$ sites and $t=t_l=1$ and $t_r=1.2$ (a) and $t=t_l=1$ and $t_r=9$ (b).}
    \label{LHN}
\end{figure}

The energy spectrum of the LHN model has one energy band which, depending on the values of $t_1$ and $t_2$ and the size of $A$, can either be a curve that is only partially complex and whose edges are real (see figure \ref{LHN} (a)) or describe a smooth closed curve on the complex plane where a winding number $\nu=\pm 1$ is well defined anywhere in its interior (see figure \ref{LHN} (b)). The spectra Figs \ref{LHN} (a) and (b) are analogous to regimes (II) and (III) of the parity-broken AL model, respectively.

\begin{figure}
    \centering
    \includegraphics[width=8.5cm]{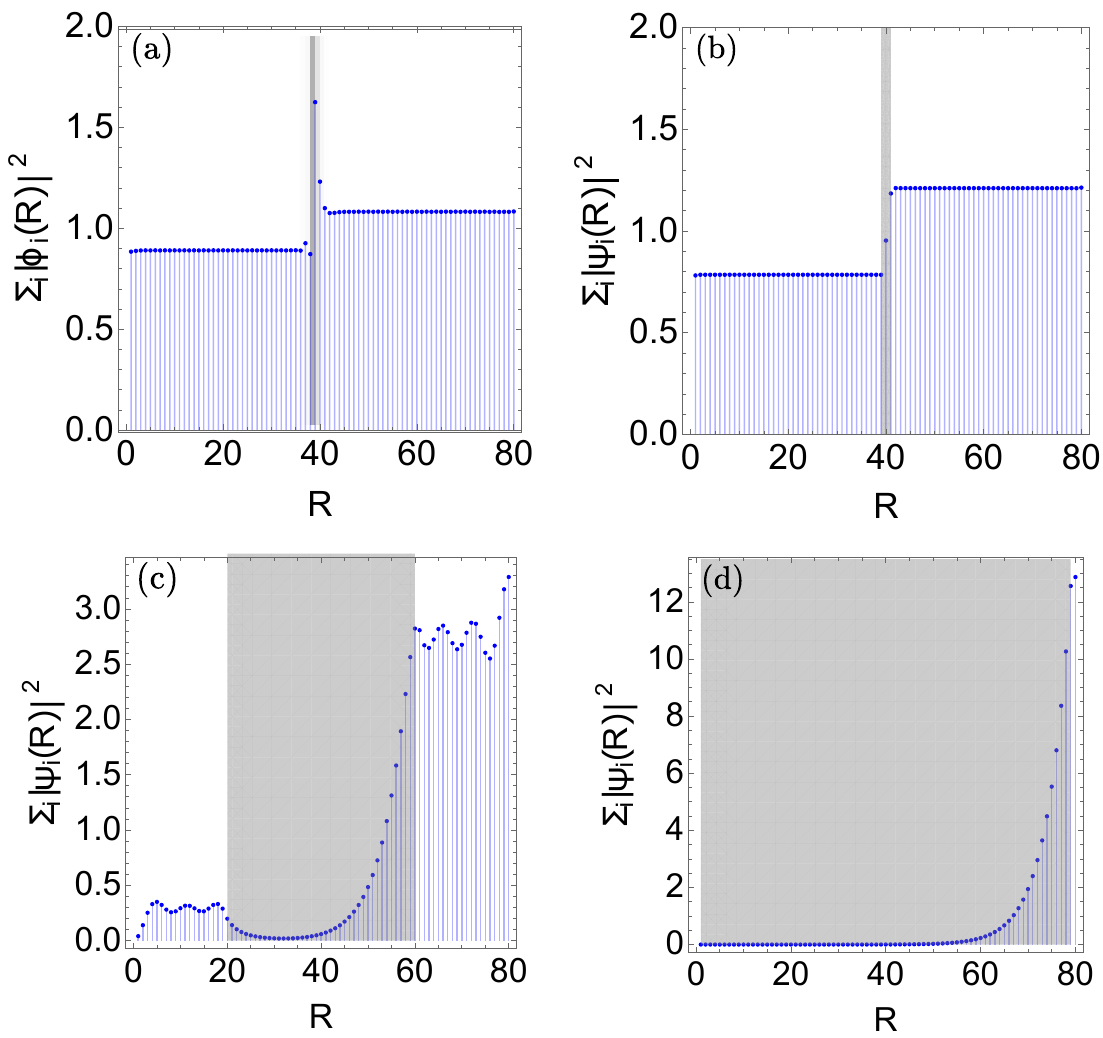}
    \caption{Local density of states with open boundary conditions of (a) a MF Hamiltonian $H[\phi]$ generated by a soliton $\phi$ at the center of the lattice and of local Hatano-Nelson Hamiltonians where the non-Hermitian defect comprises two (b), forty (c) and seventy-eight (d) sites.}
    \label{Imbalance}
\end{figure}

A translation-invariant non-reciprocal linear system with a point gap and a non-zero winding number manifests a skin effect, whereby all of the system's eigenstates accumulate at one of the boundaries. This is a direct consequence of the nonreciprocal hoppings being uniformly present throughout the entire lattice. In contrast, the nonreciprocal solitons $\phi$ in our nonlinear system spontaneously break translation invariance, causing nonreciprocal hoppings in the mean-field Hamiltonian $H[\phi]$ to exist only in the region with soliton support. 
This crucial difference is evidenced in the local density of states (LDOS) shown in Fig. \ref{Imbalance} (a) for the mean-field Hamiltonian $H[\phi]$ generated by a soliton $\phi$ located at the center of the lattice. There is an imbalance in the LDOS to the right and left sides of the soliton, albeit way less pronounced than the extreme accumulation of states at one boundary that characterizes extended non-Hermitian systems supporting the linear skin effect, such as the Hatano-Nelson (HN) model~\cite{PhysRevLett.77.570,PhysRevB.56.8651}.


To conceptually connect our model with the well-known skin effect of extended linear systems, we compare the LDOS of our model with that of Hatano-Nelson-like lattices with an increasing number of non-reciprocal hoppings. Figure \ref{Imbalance} (b) shows that the LDOS in the LHN model when the ``non-Hermitian defect'' $A$ consists of two sites present the same imbalance as in our nonlinear model with OBC. As we increase the size of $A$ until it comprises the whole system, we smoothly recover the DOS of the usual extended HN model where all states accumulate at one of the edges [see Fig.\ref{Imbalance} (c-d)].

\section{Long-time behavior of nonreciprocal solitons}

\begin{figure}
    \centering
    \includegraphics[width=0.6\linewidth]{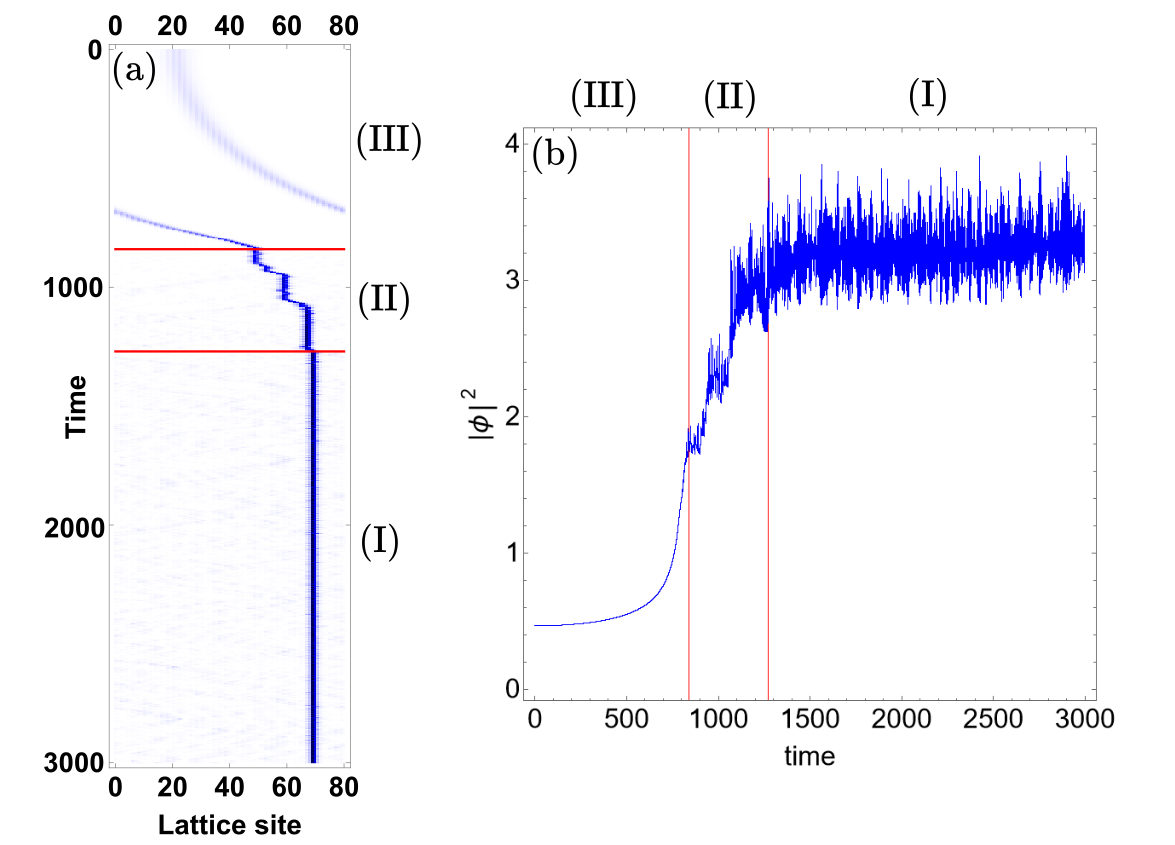}
    \caption{Dynamics of nonreciprocal solitons. (a) Time evolution of an initially static soliton with norm $|\phi|=0.68$, energy $\omega=-2.09$ and period of phase oscillation $\tau = \frac{2\pi}{|\omega|}=3.01$ from $t=0$ to $t=3000$ (roughly 997 periods of oscillation). Time runs downward and the intensity of the blue color is proportional to the probability density function of the soliton at a given site. (b) Time evolution of the intensity $|\phi^2|$ of the soliton field. In both plots, the red lines indicate the instants $t=840 \approx 279 \tau $ and $t=1270 \approx 422 \tau$.}
    \label{SteadyState}
\end{figure}

Our analysis of the three different topological regimes applies to solitons with zero instantaneous velocity. For moving solitons, the effect ``pinning down'' due to the self-focusing nature of the nonlinearity dominates at higher values of the field intensity compared to static solitons. We verified this statement by simulating the evolution of the soliton profile over long time periods, as shown in Fig. \ref{SteadyState}. The initial conditions in that simulation are the same as in Fig. (1b). Initially, the soliton smoothly accelerates, therefore being in regime (III). As the field intensity grows stronger, it transitions into the intermediate regime (II) at $t \approx 840 \approx 279 \tau$ (where $\tau=3.01$ is the period of phase oscillation of the initial soliton), when it becomes pinned down but still nonreciprocally unstable. Then, as the intensity continues to grow, the soliton falls into regime (I) at $t \approx 1270 \approx 422 \tau$, reaching a quasi-steady state where the field intensity fluctuates but does not grow much on average. We interpret this quasi-steady state as being a static soliton of regime (III) plus effectively random perturbations originating from the likely chaotic dynamics of the parity-broken AL model.

\section{The Ablowitz-Ladik equation in electrical circuits}

Any system exhibiting the AL-type nonlinear coupling is necessarily open. Topolectrical circuits~\cite{lee2018topolectrical} provide a wide variety of active and nonlinear elements, such as operational amplifiers, transistors, and varactor diodes, and thus are a straightforward platform for implementing the non-conservative, asymmetrical, and configuration-dependent hopping mechanism of the AL equation.The key idea is to engineer a circuit whose voltage dynamics obey an equation structurally equivalent to our mean-field Hamiltonian \eqref{SelfConsistent}.

We may write the Ablowitz-Ladik equation as
\begin{align}
    \mathrm{i}\frac{d}{dt} \Phi_n &= -(\epsilon+g|\Phi_n|^2)(\Phi_{n-1}+\Phi_{n+1}) = \sum_{m}H_{nm}(\Phi)\Phi_m, \nonumber \\
    H_{nm}(\Phi) &= -(\epsilon+g|\Phi_n|^2)(\delta_{n-1,m}+\delta_{n+1,m}),
\end{align}
where $\Phi_n$ are complex quantities defined at the sites $n = 1,2,\dots$ of a one-dimensional lattice, $\epsilon>0$ controls the strength of the linear nearest-neighbor (NN) hoppings from $n-1 \to n$ and $n+1 \to n$, and $g|\Phi_n|^2$ with $g \in \mathbb{R}$ now conveniently controlling the nonlinear contribution to the NN hoppings. The object $H(\Phi)$ is a solution-dependent dynamical matrix whose explicit form reads
\begin{equation}
    H(\Phi) = -\begin{pmatrix}
         0 & \epsilon + g|\Phi_1|^2 & & & \\
         \epsilon + g|\Phi_2|^2 & 0 & \epsilon + g|\Phi_2|^2 & & \\
         & \epsilon + g|\Phi_3|^2 & 0 & & \\
         & & \epsilon + g|\Phi_4|^2 & \\
         & & \ddots & \ddots
    \end{pmatrix}. \label{ALHamiltonian}
\end{equation}

We propose a realization of the dynamics generated by \eqref{ALHamiltonian} in an electrical circuit driven by an AC voltage at frequency $\Omega$, where
\begin{equation}
    V_n(t) = \Phi_n(t)e^{-\mathrm{i}\Omega t}+\Phi^{*}_n(t)e^{\mathrm{i}\Omega t} \label{Potential}
\end{equation}
is the instantaneous potential at the sites $n=1,\dots,L$ that constitute the circuit. We also assume that the envelope of $V_n(t)$ varies at a much lower rate than the driving frequency,
\begin{equation}
    \left|\frac{d}{dt}\Phi_n\right| \ll |\Omega\Phi_n|. \label{assumption}
\end{equation}

In addition to the potential, we are interested in the current $I_n$ injected by a measuring device at each node $n$ (see red arrows in Fig. \ref{Figure4}). Current conservation implies $I_n$ equals the total current flowing out of node $n$ into linked nodes $m$, which themselves may depend on the potentials $V_n$ and $V_m$ at the origin and the destination. One may express the relationship between the currents and the potentials through
\begin{equation}
    I_n = \sum_{m}L_{nm}(V) V_m,
\end{equation}
where $L(V)$ is the circuit Laplacian which, in the presence of nonlinear components, has a dependence on $V = (V_1,\dots,V_L)$. The eigenvectors and eigenvalues of the circuit Laplacian are the stationary voltage distributions and natural frequencies of the circuit. Therefore, $L(V)$ plays the same role in the dynamics of the circuit as the Hamiltonian $H(\Phi)$ does for the AL equation \eqref{equationAL}~\cite{lee2018topolectrical}.

In this context, circuit elements act as the couplings between different sites of our one-dimensional lattice. The key to reproduce the features of the AL model in an electrical circuit is to implement nonlinear capacitive couplings such that the current flowing from node $n$ to $n+1$ depends only on $V_n$ and the current flowing from node $n+1$ to $n$ depends only on $V_{n+1}$.

\subsection{Circuit design}

\begin{figure}
\begin{center}
\includegraphics[width=8cm]{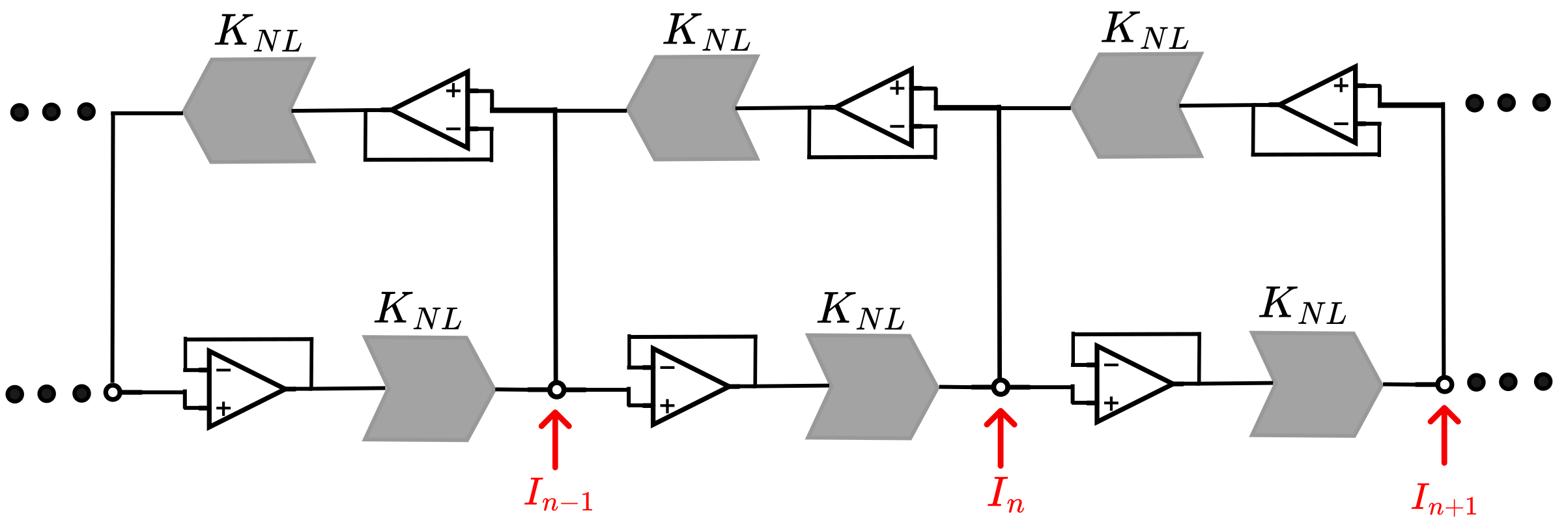}
\caption{Schematic of the topolectric circuit implementing voltage-dependent asymmetric nonlinear couplings, where $K_{NL}$ refers to the nonlinear capacitor defined in Fig. \ref{Figure3}. of the appendix }
\label{Figure4}
\end{center}
\end{figure}

The schematics of our AL circuit appear in Fig. \ref{Figure4}, where a top chain of nonlinear elements $K_{\mathrm{NL}}$ and operational ampliers (op-amps) is connected in parallel to the nodes (white circles) of a mirrored bottom chain of nonlinear elements $K_{\mathrm{NL}}$ and op-amps. The quantities we propose to directly measure are the injection currents $I_n$ at each node of the bottom chain and the effective potentials $\tilde{V}_n$ (to be defined later).

The op-amps in Fig.~\ref{Figure4} act as voltage buffers: the potentials at their input (non-inverting terminal ``+'') and output are equal, but their input current is zero. As a consequence, any circuit elements placed at the output are isolated from the current procuded by whatever load is present at the input. More concretely, the op-amps on the top (bottom) chain isolate node $n$ from the top (bottom) nonlinear element $K_{\mathrm{NL}}$ connected to node $n-1$ ($n+1$).

\begin{figure}[h!]
\begin{center}
\includegraphics[width=7cm]{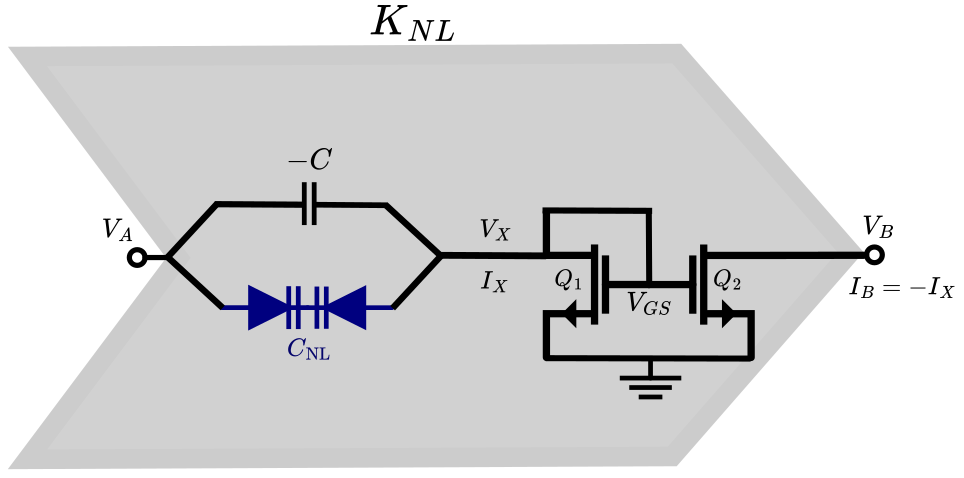}
\caption{Schematic of the nonlinear capacitor $K_{\mathrm{NL}}$ connecting nodes $A$ and $B$ whose capacitance depends on $V_A$ but not on $V_B$. This circuit employs two identical diode-connected n-type MOSFETs $Q_1$ and $Q_2$ to yield a current mirror, where $V_{\mathrm{GS}}$ represents the gate-to-source voltage.}
\label{Figure3}
\end{center}
\end{figure}

Fig. \ref{Figure3} shows the schematics of the nonlinear element $K_{\mathrm{NL}}$ with connecting nodes $A$ and $B$. There, a linear capacitor $-C$~\footnote{One can straightforwardly implement a negative capacitance by connecting $C$ in series with a negative impedance converter with current inversion (INIC) formed by an op-amp and two identical Ohmic elements -- the same configuration as employed, for instance, in \cite{PhysRevResearch.4.043108}. The output current of the INIC is the negative of the input current, thus generating an effective capacitance $-C$ in the forward direction.} and two back-to-back varactor diodes are placed in parallel. They pull a current at point $X$ given by
\begin{equation}
    I_X = (C_{\mathrm{NL}}-C)\frac{d}{dt}(V_A-V_{X}). \label{IX}
\end{equation}
The current $I_X$ then reaches a current mirror consisting of two diode-connected n-type MOSFETs $Q_1$ and $Q_2$ operating in the saturation regime, where $V_{\mathrm{GS}}$ is their gate-to-source voltage, which determines the current flowing through the transistors. The gates of both $Q_1$ and $Q_2$ are set at a potential $V_{\mathrm{GS}}$, which is connected to the drain of $Q_1$, but not to the drain of $Q_2$.

Since the drain of $Q_1$ is at point $X$, we have $V_{\mathrm{GS}}=V_X$, and all the current $I_X$ will flow from the drain of $Q_1$ to its grounded source. The second transistor $Q_2$, having the same gate-to-source voltage $V_{\mathrm{GS}}=V_{X}$ as $Q_1$, pulls a current of equal magnitude from its drain at $B$ to its grounded source, that is,
\begin{equation}
    I_B = -I_X. \label{IB}
\end{equation}

Now, let us explicitly determine $I_B$. First, we note that the capacitance $C_{\mathrm{NL}}$ composed of the back-to-back varactors is given by~\cite{10.1063/1.3418556}
\begin{equation}
    C_{\mathrm{NL}}(\Tilde{V}_A) = C_0 \ \mathrm{sech}(\alpha \Tilde{V}_A), \label{NonlinearCapacitance}
\end{equation}
where $\tilde{V}_{A} \equiv V_{A}-V_X$ is the \emph{effective potential} at node $A$, $C_0>0$ (units of capacitance) and $0<\alpha<1$ (units of voltage$^{-1}$) are constants. Then, for $|\alpha V| \ll 1$, we can approximate \eqref{NonlinearCapacitance} as
\begin{equation}
    C_{\mathrm{NL}}(\tilde{V}_A) \approx C_0 \left(1-\frac{\alpha^2}{2}\tilde{V}_A^2\right).
\end{equation}
This way, \eqref{IX} and \eqref{IB} imply
\begin{align}
    I_B &= (C-C_{\mathrm{NL}}(\tilde{V}_A)) \frac{d\Tilde{V}_A}{dt} \nonumber \\
    &= \left(\epsilon+g \tilde{V}_A^2\right)\frac{d\Tilde{V}_A}{dt},\label{IB2}
\end{align}
where we have defined $\epsilon \equiv C-C_0$ and $g \equiv C_0\alpha^2/2$. In summary, we have shown that the output current of the nonlinear element $K_{\mathrm{NL}}$ of Fig. \ref{Figure3} depends on the effective potential at the input $A$ but not on the potential at the output $B$.

The circuit as a whole is being driven at a frequency $\Omega$, so we may express the effective potential at node A as
\begin{equation}
    \Tilde{V}_{A}(t) = \Phi(t)e^{-\mathrm{i}\Omega t}+\Phi^{*}(t)e^{\mathrm{i}\Omega t},
\end{equation}
for some complex function $\Phi(t)$. The assumption \eqref{assumption} yields the following approximation for the time derivative of $\Tilde{V}_A(t)$:
\begin{equation}
    \frac{d\Tilde{V}_A}{dt} \approx \mathrm{i}\Omega \left(-\Phi e^{\mathrm{-i\Omega t}} + \Phi^{*} e^{\mathrm{i}\Omega t}\right). \label{Derivative}
\end{equation}
We also have
\begin{align}
    \Tilde{V}_A^2 = 2|\Phi|^2 + \Phi^{2}e^{-2\mathrm{i}\Omega t} + \Phi^{*2}e^{2\mathrm{i}\Omega t}, \label{Square}
\end{align}
Substituting \eqref{Derivative} and \eqref{Square} into \eqref{IB2} yields
\begin{align}
    I_B &= \mathrm{i}\Omega\left[\epsilon+g(2|\Phi|^2+\Phi^2e^{-2\mathrm{i}\Omega t}+\Phi^{*2}e^{2\mathrm{i}\Omega t})\right] \nonumber \\
    & \ \times ( -\Phi e^{-\mathrm{i}\Omega t} + \Phi^{*}e^{\mathrm{i\Omega t}}) \nonumber \\[0.3cm]
    &= -\mathrm{i}\Omega\left(\epsilon\Phi+2g|\Phi|^2\Phi-g\Phi^2\Phi^{*}\right)e^{-\mathrm{i}\Omega t} \nonumber \\ & \ + \mathrm{i}\Omega\left(\epsilon\Phi^{*}+2g|\Phi|^2\Phi^{*}-g\Phi^{*2}\Phi\right)e^{\mathrm{i}\Omega t} \nonumber \\[0.3cm]
    & \ - \mathrm{i}\Omega g\Phi^3e^{-3\mathrm{i}\Omega t}+ \mathrm{i}\Omega g\Phi^{*3}e^{3\mathrm{i}\Omega t}. \label{CurrentNonlinear}
\end{align}
Since we are only interested in the signal at frequency $\Omega$ and the third harmonics $\sim e^{\pm3\mathrm{i}\Omega t }$ can only affect the fundamental modes $\sim e^{\pm\mathrm{i}\Omega t }$ through processes of order $g^2$ or greater, and since one could alternatively filter out the third harmonics by inserting low bandpass filters in series with the circuit at each node, from now on we ignore them in \eqref{CurrentNonlinear} and write
\begin{equation}
    I_B \approx -\mathrm{i}\Omega\left(\epsilon+g|\Phi|^2\right)\Phi e^{-\mathrm{i}\Omega t} + \mathrm{i}\Omega\left(\epsilon+g|\Phi|^2\right)\Phi^{*}e^{\mathrm{i}\Omega t}.
\end{equation}

Now, looking back at Fig. \ref{Figure4}, we see that the current going from node $n-1$ to $n$ is a function of $\tilde{V}_{n-1}$, while the current flowing from node $n+1$ to $n$ is a function of $\tilde{V}_{n+1}$. By employing the conservation of current, we obtain the current being injected into node $n$ as
\begin{align}
    I_n(t) &\equiv i_n(t)e^{-\mathrm{i\Omega t}}+i_{n}^{*}(t)e^{\mathrm{i}\Omega t} \nonumber \\
    &= -\mathrm{i}\Omega\left(\epsilon+g|\Phi_{n-1}|^2\right)\Phi_{n-1} e^{-\mathrm{i}\Omega t} \nonumber \\
    & \ + \mathrm{i}\Omega\left(\epsilon+g|\Phi_{n-1}|^2\right)\Phi_{n-1}^{*}e^{\mathrm{i}\Omega t} \nonumber \\
    & \ -\mathrm{i}\Omega\left(\epsilon+g|\Phi_{n+1}|^2\right)\Phi_{n
    +1} e^{-\mathrm{i}\Omega t} \nonumber \\ 
    & \ + \mathrm{i}\Omega\left(\epsilon+g|\Phi_{n+1}|^2\right)\Phi_{n+1}^{*}e^{\mathrm{i}\Omega t}. \label{Expression2}
\end{align}
By collecting the terms proportional to $e^{\mathrm{i}\Omega t}$ and $e^{-\mathrm{i}\Omega t}$, we can re-express \eqref{Expression2} as
\begin{align}
    i_n(t) &= -\mathrm{i}\Omega\left[\epsilon(\Phi_{n-1}+\Phi_{n-1})+g|\Phi_{n+1}|^2\Phi_{n+1}\right] \nonumber \\
    &\equiv \mathrm{i}\Omega \sum_{m}H_{nm}(\Phi)\Phi_m\nonumber \\
    i_n^{*}(t) &= \mathrm{i}\Omega\left[\epsilon(\Phi^{*}_{n-1}+\Phi^{*}_{n-1})+g|\Phi_{n+1}|^2\Phi^{*}_{n+1}\right] \nonumber \\
    &\equiv -\mathrm{i}\Omega \sum_{m}H_{nm}(\Phi)\Phi_m^{*} \label{In2}
\end{align}

The explicit form of the matrix $H(\Phi)$ in \eqref{In2} is
\begin{equation}
    H = -\begin{pmatrix}
         0 & \epsilon + g|\Phi_2|^2 & & & \\
         \epsilon + g|\Phi_1|^2 & 0 & \epsilon + g|\Phi_3|^2 & & \\
         & \epsilon + g|\Phi_2|^2 & 0 & & \\
         & & \epsilon + g|\Phi_3|^2 & 0 & \\
         & & \ddots & \ddots 
    \end{pmatrix}, \label{Transpose}
\end{equation}
which is just the transpose of the AL mean-field Hamiltonian \eqref{ALHamiltonian}. Since we haven't broken parity symmetry yet, the matrices \eqref{ALHamiltonian} and \eqref{Transpose} generate the same (reciprocal) dynamics. 

One may brake parity in the circuit by alternating the strength of the linear capacitors as $C \to C_1,C_2$ (which is analog to having dimerized linear couplings $t_1$ and $t_2$), and by connecting elements with staggered conductances to each node (which is analog to adding on-site energies $\pm\Delta$). Such elements can be linear capacitors or inductors. This way, the resulting circuit Laplacian will be equivalent to the transpose of our parity-broken AL mean-field Hamiltonian in Eq. 5 of the manuscript, and thus would describe the evolution of a parity-broken AL system with nonreciprocity in the opposite direction, that is, with identical but mirrored dynamics.

\bibliography{main}